\documentclass[aps,pra,reprint,groupedaddress]{revtex4-2}
\usepackage{graphicx}  
\usepackage{amsmath,amssymb}
\usepackage{xcolor}
\usepackage[normalem]{ulem}
\usepackage{hyperref}

\DeclareMathOperator*{\SumInt}{%
\mathchoice%
  {\ooalign{$\displaystyle\sum$\cr\hidewidth$\displaystyle\int$\hidewidth\cr}}
  {\ooalign{\raisebox{.14\height}{\scalebox{.7}{$\textstyle\sum$}}\cr\hidewidth$\textstyle\int$\hidewidth\cr}}
  {\ooalign{\raisebox{.2\height}{\scalebox{.6}{$\scriptstyle\sum$}}\cr$\scriptstyle\int$\cr}}
  {\ooalign{\raisebox{.2\height}{\scalebox{.6}{$\scriptstyle\sum$}}\cr$\scriptstyle\int$\cr}}
}

\begin{document}
\title{Two-color polarization control on angularly resolved attosecond time delays}

\author{D.I.R. Boll}
\email[]{boll@ifir-conicet.gov.ar}
\affiliation{Instituto de Física Rosario, CONICET-UNR, Ocampo y Esmeralda, 2000 Rosario, Argentina}

\author{L. Martini}
\affiliation{Instituto de Física Rosario, CONICET-UNR, Ocampo y Esmeralda, 2000 Rosario, Argentina}
\affiliation{Department of Physics and Astronomy, University of Southern California, Los Angeles, California 90089-0484, USA}

\author{A. Palacios}
\affiliation{Departamento de Química, Módulo 13, Universidad Autónoma de Madrid, 28049 Madrid, Spain}
\affiliation{Institute of Advanced Research in Chemical Sciences (IAdChem), Universidad Autónoma de Madrid, 28049 Madrid, Spain}

\author{O.A. Foj\'on}
\affiliation{Instituto de Física Rosario, CONICET-UNR, Ocampo y Esmeralda, 2000 Rosario, Argentina}
\affiliation{Escuela de Ciencias Exactas y Naturales, FCEIA, Universidad Nacional de Rosario, Argentina}

\date{\today}

\begin{abstract}
Measured photoionization time delays may exhibit large variations as a function of the emission angles, even for spherically symmetric targets, as shown in recent RABBITT (reconstruction of attosecond beating by interference of two-photon transitions) experiments. The contributions from different pathways to the two-photon quantum channels can already explain the observed phase jumps that shape those angular distributions. Here, we propose a simple analytical model to describe angularly-resolved RABBITT spectra as a function of the relative polarization angle between the ionizing attosecond pulse train and the assisting IR field. We demonstrate that the angular dependencies of the measured delays can be analytically predicted and the position of the phase jumps reduced to the analysis of a few relevant parameters. 
     
\end{abstract}

\maketitle

\section{Introduction\label{sec:intro}}
The chronoscopy of electron photoemission \cite{Pazourek2015} turned into reality upon the mastery of two pump-probe attosecond spectroscopy techniques: the attosecond streaking \cite{Hentschel2001} and the RABBITT \cite{Veniard1996,Paul2001}(reconstruction of attosecond beating by interference of two-photon transitions). While born as characterization procedures for attosecond pulses, they become the most successful experimental approaches to access the so-called Eisenbud-Wigner-Smith (EWS) time delays for single-photon ionization processes \cite{Pazourek2015,deCarvalho2002,Cavalieri2007,Schultze2010,Klunder2011}. The EWS photoionization time delay holds information on the initial state and the potential felt by the electron while escaping. Thus, it conveys rich structural and dynamical information on the potential landscape where electrons evolve. Accordingly, the EWS time delay will depend on photoelectron energy and its emission direction unless a single partial wave in the continuum prevails \cite{Pazourek2015,deCarvalho2002,Froissart1963,Pazourek2013}.

The retrieval of EWS time delays from attosecond spectroscopic experiments requires a careful examination of the effects induced by the probe field \cite{Nagele2011,Dahlstrom2013}. The frequency spectrum of typical attosecond pulse trains in RABBITT experiments comprises a comb of extreme ultraviolet (XUV) odd-order harmonics from a fundamental infrared (IR) laser, giving rise to photoelectron mainbands. Consequently, the ionization of a target with an attosecond pulse train assisted with a phase-locked IR, leads to a another set of photoelectron spectral lines (sidebands), which lie in between the mainbands and correspond to the interfering two-photon (XUV $\pm$ IR) channels. This interference leads to a pump-probe delay $(\tau)$ dependent signal in the sideband that follows the expression \cite{Paul2001}  
\begin{align}\label{eq:SB-signal-fit}
I_{2q}(\tau) \propto A+B\cos(2\omega_0\tau - \phi_{at}),
\end{align}where $\omega_0$ is the IR photon frequency, and $\phi_{at}\,(=2\omega_0 \tau_{at})$ is the phase difference that carries the total photoelectron emission delay $(\tau_{at})$, which results from the EWS and the measurement-induced continuum-continuum (cc) time delays \cite{Dahlstrom2013}.

Previous studies established that total atomic delays usually follow the additive relation $\tau_{at}=\tau_{EWS}+\tau_{cc}$ for atomic targets in non-resonant angle-integrated measurements \cite{Klunder2011,Palatchi2014,Guenot2014}. However, the lack of spherical symmetry prevents finding a direct connection between total and EWS time delays in molecular targets \cite{Huppert2016,Baykusheva2017}. The situation is much more challenging in angle-resolved measurements \cite{Heuser2016,Cirelli2018,Busto2019,Autuori2022,Vos2018,Nandi2020,Holzmeier2021,Ahmadi2022}. In that case, angular dependencies induced on the IR probing stage may be the primary source of measured time delay anisotropies. Indeed, for initial $s$ states in atomic systems, the EWS time delay is strictly angle-independent, whereas the total time delays retrieved from RABBITT experiments display steep variations for some emission angles \cite{Heuser2016}. These angular variations, attributed to small phase-differences induced on the cc transitions \cite{Heuser2016}, are also explained by destructive interferences on the absorption channel, governed by the relative weight of partial waves populated by the IR field \cite{Busto2019}.

To our knowledge, there is no previously published analytical models that can predict the angular dependencies induced by cc transitions on RABBITT experiments. Here, we show for the first time that a simplified analytic model for the two-photon matrix elements can accurately describe the angular variations of time delays, particularly those induced by the IR probe. To that end, we consider the total time delay for initial $s$ states in atomic systems and compare it with available experimental data \cite{Heuser2016}. Then, we resort to the polarization control technique \cite{OKeeffe2004,Meyer2008} to externally manipulate the relative weight of partial waves. In that case, the angular evolution of time delays displays marked differences \cite{Jiang2022}. We show that a new interference mechanism is co-responsible for the observed total time delay variations in that scenario.

\section{Theoretical methods}\label{sec:theory}
The angular variation of atomic time delays is intrinsically linked with the photoelectron angular distributions (PADs) \cite{Heuser2016,Ivanov2017,Hockett2017,Busto2019,Fuchs2020}. The most general expression for PADs is given by \cite{Reid2003}
\begin{align}\label{eq:SB-AR}
I_{2q}(\theta,\phi)\propto \Biggl \vert \sum_{L,M} M_{L,M} Y_L^M(\theta,\phi)\Biggr \vert^2,
\end{align}where $\theta$ and $\phi$ are the angles defining the electron emission direction, and $Y_L^M$ is a spherical harmonic function. The method of partial waves, implicit in the equation above, allows us to obtain a convenient expression of the transition matrix amplitudes $M_{L,M}$ \cite{Joachain1975} that connect the initial bound state of the atomic target with a final continuum state, labeled by the azimuthal and magnetic quantum numbers $L$  and $M$.

In the RABBITT spectra, the time delay information emerges in the sidebands \cite{Klunder2011,Dahlstrom2013}. The quantum pathways involving the absorption $(+)$ or emission $(-)$ of an IR photon from consecutive odd-harmonics add coherently. The transition matrix element can be then formally separated into two terms
\begin{align}\label{eq:M-split}
M_{L,M}=M_{L,M}^+ + M_{L,M}^-,
\end{align}that collect the contributions to each channel. For typical IR intensities in RABBITT experiments, the analysis of $M_{L,M}^{\pm}$ can be carried out through second-order perturbation theory (SOPT). In the limit of infinitely long pulses, each transition matrix amplitude reads
\begin{align}\label{eq:TMA}
M_{L,M}^{\pm}= e^{\pm i \omega_0 \tau} \sum_{\lambda,\mu} A_{L,\lambda,l_i}^{M,\mu,m_i} T_{L,\lambda}^{l_i,\pm}. 
\end{align} The summation above runs over the azimuthal $(\lambda)$ and magnetic $(\mu)$ quantum numbers of intermediate states allowed by the application of the electric dipole selection rule from the initial state with angular quantum numbers $l_i$ and $m_i$. The angular factors $A_{L,\lambda,l_i}^{M,\mu,m_i}$ are identical for absorption and emission channels, but the (pseudo) radial matrix elements $T_{L,\lambda}^{l_i,\pm}$, in general, are not \cite{Heuser2016,Busto2019,Boll2022a}. 

\begin{figure}
 \includegraphics[width=\columnwidth,clip]{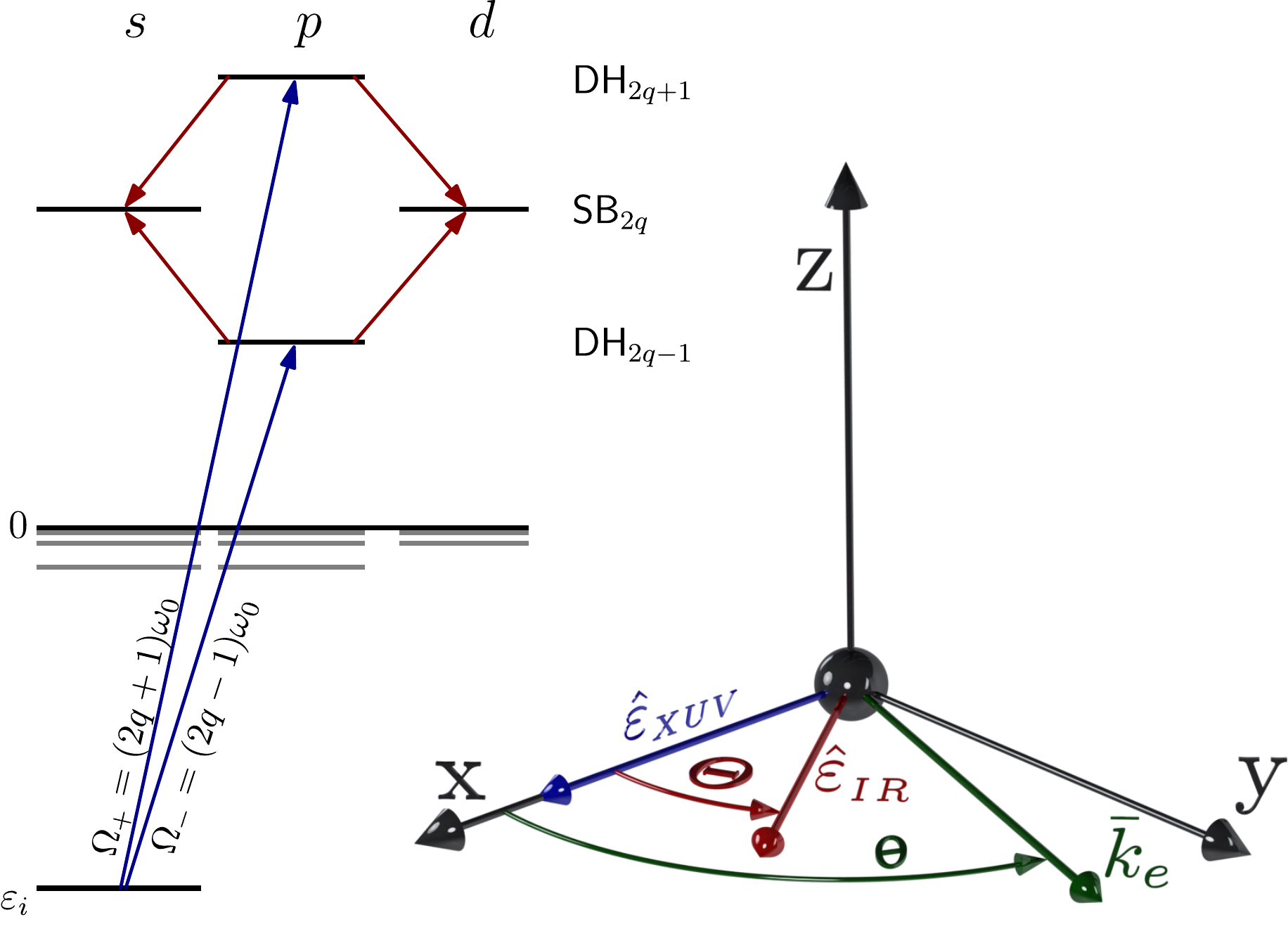}%
 \caption{The left panel shows a schematic representation of pathways contributing to consecutive mainbands and the sideband in-between. For low IR intensities, mainbands are mainly populated by transitions from ground state to continuum states, triggered by the absorption of a single photon with energy $\Omega_{\pm}=(2q\pm1)\omega_0$, from the attosecond pulse train. The interaction of these primary photoelectron distributions with the IR gives rise to the sidebands. The reverse contributions, where the IR photon is absorbed or emitted first, are not shown here. The right panel shows the geometric configuration. We indicate the polarization direction of the XUV laser source by the blue vector $\hat{\varepsilon}_{XUV}$ collinear to the cartesian $x$ axis. The polarization vector of the IR laser $\hat{\varepsilon}_{IR}$ lies in the $xy$ plane, and subtends an angle $\Theta$ with the cartesian $x$ axis. The final photoelectron momentum also lies in the $xy$ plane, and it is indicated by the green vector $\mathbf{k}_e$. \label{fig:scheme}}
\end{figure}

In the two-color case, the latter splits into the sum of two terms
\begin{align}\label{eq:TL-split}
T_{L,\lambda}^{l_i,\pm}=T_{L,\lambda}^{l_i}(\Omega_{\mp}) + T_{L,\lambda}^{l_i}(\pm\omega_0),
\end{align} describing the possible time orderings for the two-photon processes that lead to the same final continuum state. The first term accounts for the absorption of one XUV photon with frequency $\Omega_{\mp}=(2q\mp1)\omega_0$ from the attosecond pulse train, and the subsequent absorption $(+)$ or emission $(-)$ of one IR photon. The other term describes the reverse process where the exchange of one IR photon is followed by the absorption of the XUV photon. In the length gauge, each term on the right-hand side of Eq. \eqref{eq:TL-split} reads
\begin{align}\label{eq:RMEs}
T_{L,\lambda}^{l_i}(\Omega)=(-i)^L e^{i\sigma_L}\SumInt \frac{\langle R_{\varepsilon_{k_e},L}\vert r \vert R_{\varepsilon_{\kappa},\lambda} \rangle \langle R_{\varepsilon_{\kappa},\lambda} \vert r \vert R_{\varepsilon_i,l_i}\rangle}{\varepsilon_i+\Omega-\varepsilon_{\kappa}},
\end{align} where $\sigma_L=\arg{\Gamma(L+1-i/k_e)}$ is the Coulomb phase for a final state with angular momentum $L$ and momentum $k_e$. The summation (integration) runs over the entire spectrum of unperturbed states of the target, with energy $\varepsilon_{\kappa}$ and radial wavefunctions $R_{\varepsilon_{\kappa},\lambda}$. The initial and final states of the system are also described by eigenstates of the field-free hamiltonian, with radial wavefunction $R_{\varepsilon_i,l_i}$ and $R_{\varepsilon_{k_e},L}$, respectively \cite{Boll2022a}.

To highlight the influence of the cc transitions on the angular variation of atomic time delays, we focus on initial $s\,(l_i=0)$ states. In turn, this initial state configuration will allow us to directly compare the results of our model with benchmark experimental and theoretical data \cite{Heuser2016,Ivanov2017,Busto2019}. The left panel in Fig. \ref{fig:scheme} schematically displays the most relevant reaction pathways contributing to a generic sideband of order $2q$. As the angular momentum of every intermediate state is $\lambda=1$, we can drop the indices $l_i$ and $\lambda$ for ease of notation. Substituing Eq. \eqref{eq:M-split} into Eq. \eqref{eq:SB-AR} and expanding it with the use of Eq. \eqref{eq:TMA}, we obtain a series whose terms can be classified into two categories based on their explicit dependence on $\omega_0 \tau$. 

On the one hand, non-oscillatory terms are independent of $\tau$, and we can group them into a variable customarily termed $A$ (or $\alpha$) \cite{Dahlstrom2013,Ivanov2017}. On the other hand, we may combine the oscillatory terms into the double series
\begin{align}\label{eq:oscillatory-terms}
\sum_{L,L'} \vert T_{L}^{+}\vert \vert T_{L'}^{-}\vert g_{L,L'} \cos(2\omega_0\tau+\phi_L^+-\phi_{L'}^-)
\end{align} where $\vert T_{L}^{\pm}\vert$ and $\phi_L^{\pm}$ are the moduli and phase of the (pseudo) radial matrix element in Eq. \eqref{eq:TL-split}, for absorption and emission pathways. The function $g_{L,L'}$ contains the full angular information. To obtain the explicit expressions for $g_{L,L'}$ (see appendix \ref{sec:appendix}) we adopt the geometric layout presented in the right panel of Fig. \ref{fig:scheme}. We choose the quantization axis for angular momentum in the $z$-direction, whereas the attosecond pulse train is linearly polarized (blue vector) along the $x$ axis. The polarization vector of the IR laser (red vector) and the photoelectron emission direction (green vector) lie in the $xy$ plane subtending the angles $\Theta$ and $\theta$ with respect to the $x$ axis, respectively. 

The series above can be summed to give $B\cos(2\omega_0\tau-\phi_{at})$ by virtue of the harmonic addition theorem \cite{HAT}. Thus, the photoelectron angular distributions of sideband lines reduce to Eq. \eqref{eq:SB-signal-fit} also for angle-resolved setups. The angular dependence is entirely contained in factors $A$, $B$, and the atomic phase $\phi_{at}$ that satisfies the equation 
\begin{align}\label{eq:trigo-phase}
\tan (\phi_{at}) = \frac{\sum_{L,L'} \vert T_{L}^{+}\vert \vert T_{L'}^{-}\vert g_{L,L'} \sin(\phi_{L'}^{-}-\phi_{L}^{+})}{\sum_{L,L'}\vert T_{L}^{+}\vert \vert T_{L'}^{-}\vert g_{L,L'} \cos (\phi_{L'}^{-}-\phi_{L}^{+})}.
\end{align}This atomic phase $\phi_{at}$ is connected to the so-called total atomic delay by $\tau_{at}=\phi_{at}/2\omega_0$ \cite{Heuser2016}, because we consider Fourier limited (chirp-free) attosecond pulses in the train \cite{Ivanov2017}.    
 
From Eq. \eqref{eq:trigo-phase}, we can unambiguously identify a necessary condition to observe angular variations in time delays: the phase of the (pseudo) radial matrix elements must depend on the angular quantum number of the final state. Otherwise, if $\phi_L^{\pm}=\phi^{\pm}$, the trigonometric functions on the right-hand side of Eq. \eqref{eq:trigo-phase} can be factorized, the contribution of the $g_{L,L’}$ functions cancels identically, and the angular dependence in $\phi_{at}$ is completely lost. 

One main contribution of this paper is to show that a purely analytic method can quantitatively describe the angular variation of atomic time delays, even for multielectronic targets. Starting from Eq. \eqref{eq:trigo-phase}, and based on the condition established in the preceding paragraph, we must resort to a model capturing the dependence of radial matrix elements on the angular quantum numbers of final states. To the best of our knowledge, the analytic continuum-continuum radial matrix elements (ACC-RME) model \cite{Boll2022a} is the only one that meets these requirements. Another appealing aspect of this model is that it demonstrates a factorization of radial matrix elements as 
\begin{align}\label{eq:TL-factor}
T_L^{\pm}=\vert T_L^{\pm}\vert e^{i\phi_L^{\pm}}\simeq \vert T_{L,bc}^{\pm}\vert \vert T_{L,cc}^{\pm}\vert e^{i(\phi_{bc}^{\pm}+\phi_{L,cc}^{\pm})},
\end{align}where $T_{L,bc}^{\pm}$ describes the ionization step of the reaction and $T_{L,cc}^{\pm}$ accounts for transitions upon exchange of one IR photon. The latter depends on the angular quantum number of final states, in contrast with previous models \cite{Dahlstrom2013}, which allows us to survey angular variations in time delays from a fully analytic perspective. In addition, the ACC-RME model satisfactorily recovers the universal character observed for the phases and moduli of cc transitions \cite{Busto2019,Fuchs2020}. Thus, the study of angularly resolved atomic time delays for initial $s$ states can be constrained to hydrogen targets without loss of generality. We further test the validity and accuracy of the model by comparing it with available experimental data, and the results from the numerical solution of the time-dependent Schrodinger Equation (TDSE), as well as the widely employed SOPT.

\section{Results}\label{sec:results}
We obtain the angle-resolved RABBITT spectra by solving the TDSE with the \emph{Qprop} code \cite{Tulsky2020}, for a comb of high-order odd harmonics and different pump-probe delays $\tau$. We extract the total atomic time delay by fitting to Eq. \eqref{eq:SB-signal-fit} the sideband signal dependence on $\tau$. Additionally, we replace into Eq. \eqref{eq:trigo-phase} the (pseudo) radial matrix elements we obtain from SOPT \cite{Jayadevan2001} and ACC-RME \cite{Boll2022a} methods. Firstly, we revisit the case of colinear polarization of pump-and-probe fields for which theoretical and experimental benchmark data is available \cite{Heuser2016,Ivanov2017,Busto2019}. Then, we explore the non-collinear case. Datasets for the atomic time delays we obtain through all these methods and for each configuration are publicly available \cite{DataBoll2022b}. Atomic units are used unless stated otherwise.

\subsection{The parallel case, $\Theta=0^{\circ}$}

\begin{figure*}
 \includegraphics[width=0.49\textwidth,clip]{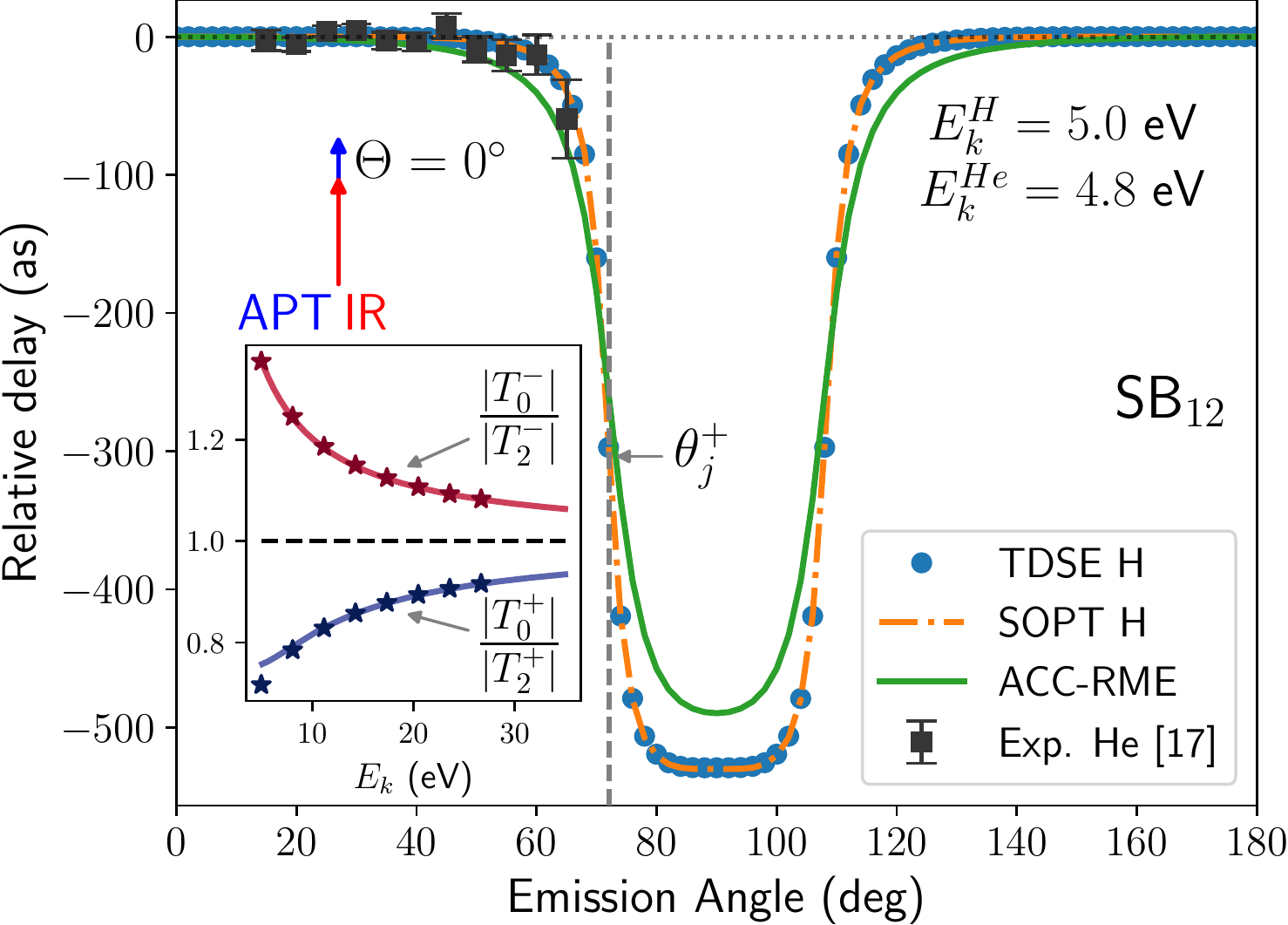}
 \includegraphics[width=0.49\textwidth,clip]{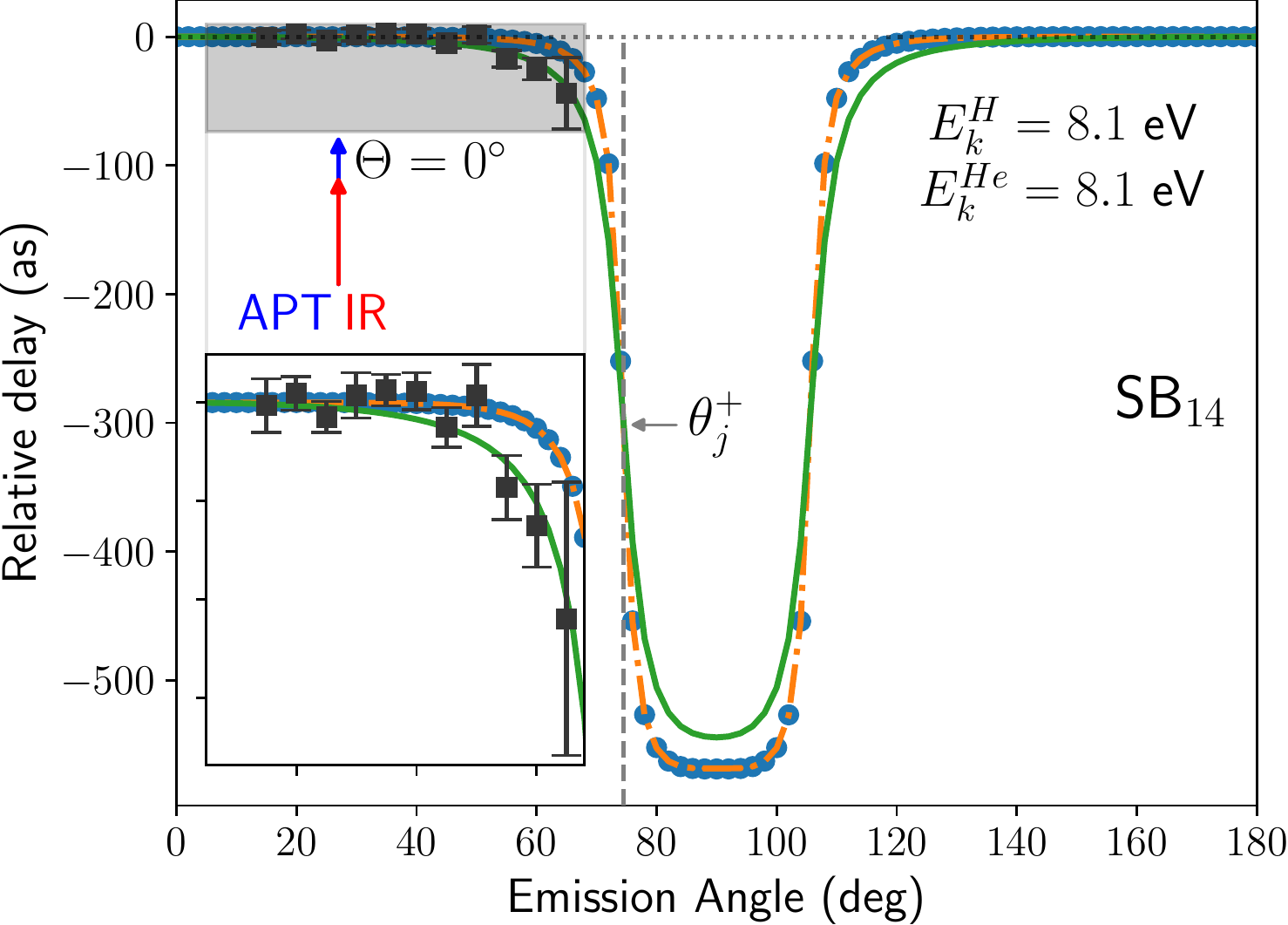}\\
 \includegraphics[width=0.49\textwidth,clip]{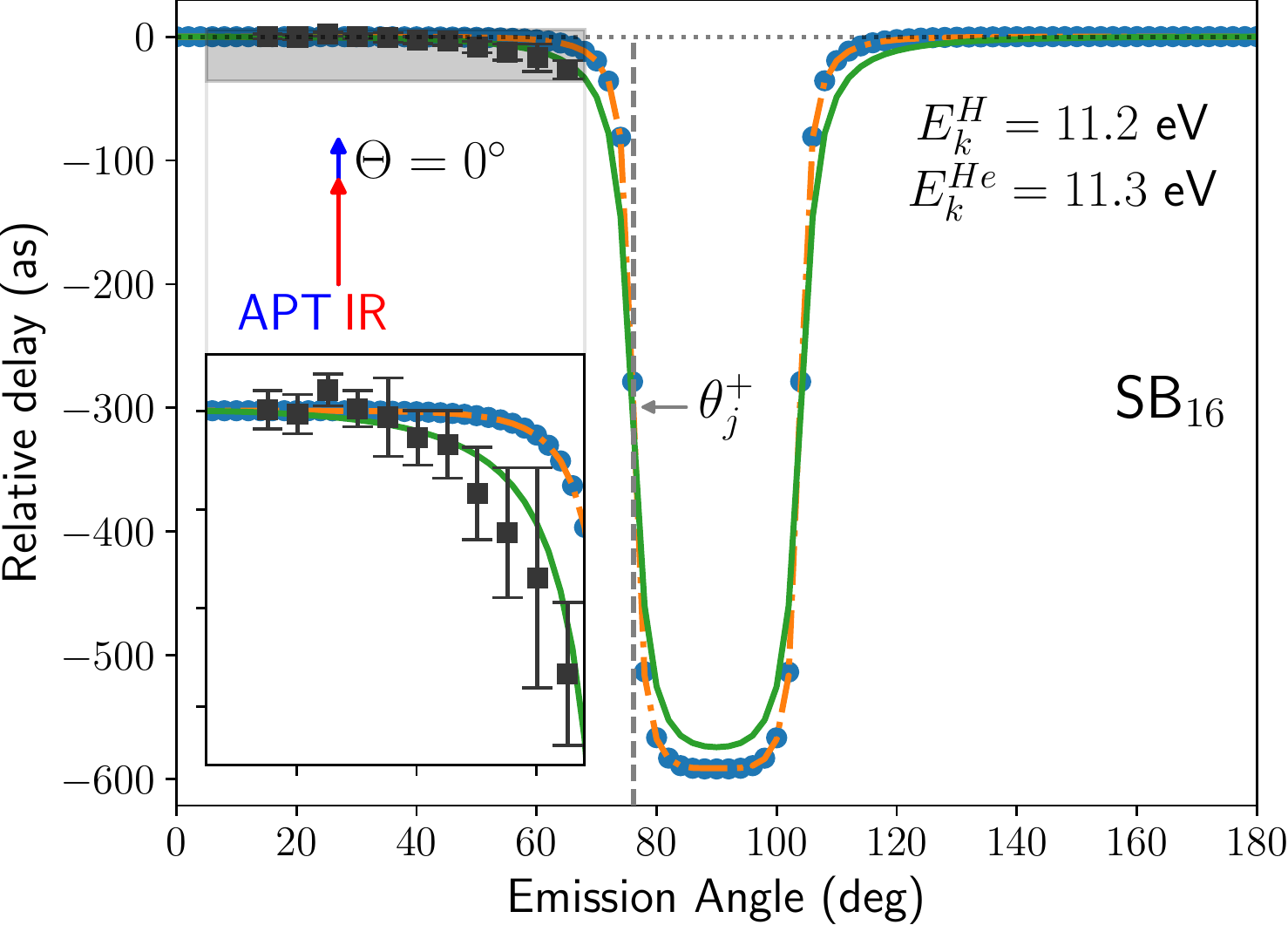}           
 \includegraphics[width=0.49\textwidth,clip]{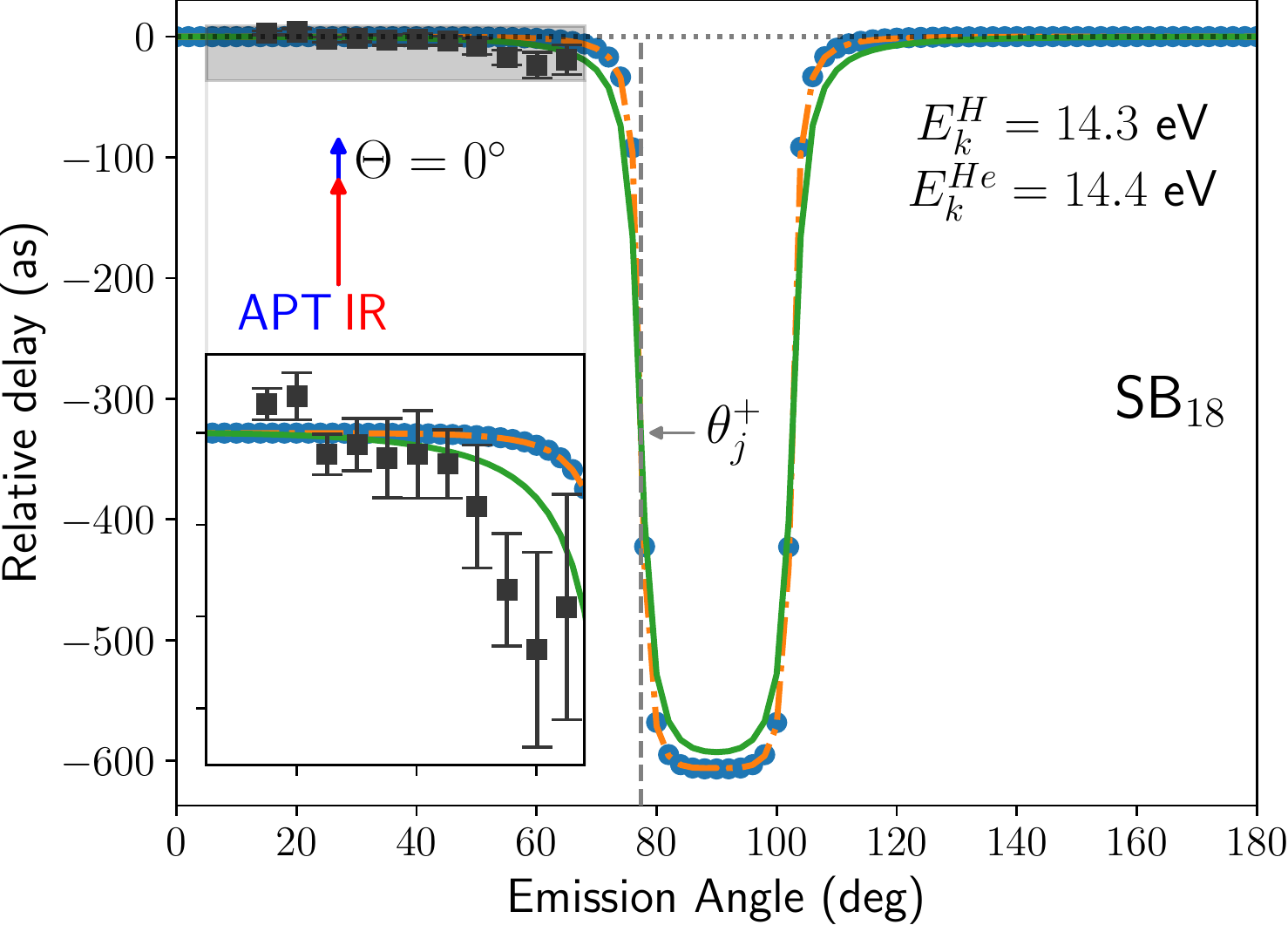}
 \caption{Comparison of the angular variation of time delays at photoelectron energies coinciding with RABBITT sidebands for hydrogen and helium atoms. Theoretical calculations for hydrogen atom are computed through three different methods: TDSE, SOPT and ACC-RME calculations. The experimental data \cite{Heuser2016} corresponds to helium atom. Photoelectron kinetic energy of final states in both systems are indicated in each panel. The sideband index corresponds to hydrogen. The phase jump angle $\theta_j^+$ is calculated from Eq. \eqref{eq:angle-phase-jump}, and the relative time delay for that emission angle is obtained from linear interpolation from the TDSE results. This leads to larger fluctuations for sidebands with larger index due to the steeper variations of $\Delta \tau_{tot}$. The inset in figure for SB$_{12}$ shows the ratios $\vert T_0^{\pm}/T_2^{\pm}\vert$ obtained from SOPT (stars) and ACC-RME (lines) calculations \cite{Boll2022a}. The other insets show the zoomed image of relative time delays for the gray-shaded areas. \label{fig:AR-parallel}}
\end{figure*}

From the theoretical point of view, we can provide precise values for the photoemission time delay. However, the experimental determination of this quantity employs attosecond pulse trains, usually generated with a frequency chirp that is difficult to retrieve. Thus, it is common to extract relative time delays with respect to a given reference. This is straightforward for angularly resolved measurements. Since there is a perfect cancellation of the XUV phase when using as a reference the signal retrieved for a specific angle. 

In Fig. \ref{fig:AR-parallel}, we present the angularly resolved total time delays for the hydrogen atom in the standard RABBITT scheme, \emph{i.e.} with parallel linearly polarized XUV attosecond pulse train and IR field. We display the relative time delays considering as reference the value for photoelectron emission parallel to the light polarization direction, $\Delta \tau_{at}(\theta)=\tau_{at}(\theta)-\tau_{at}(0^{\circ})$. The results correspond to sidebands 12 to 18 in hydrogen, initially in its ground state. Model predictions (green full line) agree with \emph{ab initio} results obtained within the SOPT (orange dashed line) and by solving the TDSE (blue circles). We observe that the mean-squared error of model results with respect to those from \emph{ab initio} calculations decreases as the photoelectron kinetic energy increases. Slight inaccuracies in the phase of each radial matrix element derived from the ACC-RME model explain the differences with the \emph{ab-initio} calculations. In contrast, the moduli (quotient) of ACC-RME radial matrix elements are in almost perfect agreement with state-of-the-art calculations, as the inset in the figure for SB$_{12}$ reveals. For comparison, we also include the experimental data for the He atom from Ref. \cite{Heuser2016}. It is interesting to note that, except for sideband 12, ACC-RME results better describe the experimental data (see insets for sidebands 14 to 18). To the best of our knowledge, this is a fortuitous circumstance. The model provides slightly larger (smaller) phase differences between partial waves $s$ and $d$ in the absorption (emission) channels, as compared to accurate calculations \cite{Boll2022a}. In turn, these shifts seem to reproduce the spectral contamination effect that pervades the experimental data \cite{Heuser2016}.

The relative time delay presents steep variations for electron emission angles $\theta$ above $65^{\circ}$, as discussed in previous studies \cite{Heuser2016,Ivanov2017,Busto2019}. In the past, the analysis of this trend focused on the spherical harmonics describing the angular dependency of photoemission \cite{Heuser2016} and their interplay with the transition matrix amplitudes \cite{Busto2019}. Here, we present an alternative procedure for studying the phase jumps in atomic time delay using simple analytical expression. We start from the definition of relative time delays $\Delta \tau_{at}(\theta)$ and combine it with the expression for atomic phases in Eq. \eqref{eq:trigo-phase}. The angular dependence of relative time delays reduces to
\begin{align}\label{eq:AR-phase}
\Delta \tau_{tot}(\theta)=\frac{1}{2\omega_0}\arctan\left[\frac{f(\theta)-f(0)}{1+f(\theta)f(0)}\right],
\end{align}where $f(\theta)$ is the right-hand side of Eq. \eqref{eq:trigo-phase}. Phase jumps will occur for zero values of the denominator in the arctan function argument. However, in order to reach a simplified analytical expression to predict the phase jumps, we will consider the very likely scenario in which the phase difference between different partial waves are almost equal for absorption and stimulated emission \cite{Fuchs2020,Boll2022a}, or equivalently:  $\phi_0^{\pm}\simeq \phi_2^{\pm}+\Delta$, with $ \vert \Delta \vert \ll 1$. Applying this approximation, and for $\Delta=0$ \footnote{The next order term in the Taylor expansion around $\Delta=0$ of the solutions to the quadratic equation are negligible due to $ \vert \Delta \vert \ll 1$}, one obtains a quadratic equation in $\cos(2\theta)$ for the phase jump angles that satisfies the following relation
\begin{align}\label{eq:angle-phase-jump}
\cos(2\theta_{j}^{\pm}) \simeq - \left[\frac{1}{3} + \frac{2}{3}\frac{\vert T_0^{\pm}\vert}{\vert T_2^{\pm}\vert}\right].
\end{align}Here, we want to remark that $\Delta \neq 0$ is a necessary condition for the obtention of atomic time delays depending on the photoelectron emission angle. Otherwise, $f(\theta)\equiv f(0)$ in Eq. \eqref{eq:AR-phase}, and no phase jump can occur because the angular dependence is lost. However, the specific values of $\Delta$ do not influence the position of phase jump angles $\theta_{j}^{\pm}$, as long as $\vert \Delta \vert \ll 1$. 

From Eq. \eqref{eq:angle-phase-jump}, we see that the moduli quotient of radial matrix elements $T_L^{\pm}$ largely dictates the photoelectron emission angles where phase jumps take place. Hence, the usefulness of Eq. \eqref{eq:angle-phase-jump} improves by combining it with the generalized Fano’s propensity rule \cite{Busto2019}. It states that cc transitions triggered by the absorption (emission) of one IR photon favor the partial wave with higher (lower) angular momentum. Thus, phase jumps shown in Fig. \ref{fig:AR-parallel} can only result from absorption pathways, as the quotient satisfies $0<\vert T_0^+ / T_2^+\vert \leq 1$,  and we can always find a real solution $\theta_j^+$ to Eq. \eqref{eq:angle-phase-jump}. The vertical dashed lines in Fig. \ref{fig:AR-parallel} display the phase jump angles $\theta_j^+$ we obtain by using radial matrix elements from SOPT calculations \cite{Jayadevan2001,Boll2022a}. Conversely, emission channels cannot induce phase jumps on this setup. The inset in Fig. \ref{fig:AR-parallel} for SB$_{12}$ quantitatively depicts the generalized Fano's propensity rule for initial $s$ states. The almost perfect agreement between exact SOPT (stars) and ACC-RME model (solid lines) results on the entire energy range illustrates the accuracy of the analytic method. Consequently, we can safely invoke the factorization of radial matrix elements that the ACC-RME model predicts to reach an additional conclusion. The phase jump angle $\theta_j^+$ derived from Eq. \eqref{eq:angle-phase-jump} will only depend on the relative strength of the continuum-continuum transitions $p\rightarrow s$ and $p \rightarrow d$. Ultimately, this allows us to conjecture a direct dependence of the phase jump angles on the photoelectron energy, the net charge of the parent ion, and the IR photon wavelength.

For comparison, in Fig. \ref{fig:AR-parallel}, we also include the experimental data for the relative total atomic delays in helium reported in Ref. \cite{Heuser2016}. The similarity between the theoretical results for hydrogen and the experimental data for helium is not surprising and follows from two facts. First, single-electron dynamics dominate the response of helium atoms in this energy range \cite{Heuser2016,Boll2020}. Second, the quotients of transition matrix amplitudes for $s$ and $d$ partial waves, $T_0^{\pm}/T_2^{\pm}$, govern the relative time delay patterns \cite{Heuser2016}. Moreover, as the moduli and phases of these quotients follow a universal pattern for the same final photoelectron energy \cite{Busto2019,Fuchs2020,Boll2022a}, the angular variation of time delays for hydrogen and helium atoms must be similar. This universal trend is also captured in the results plotted in Fig. \ref{fig:AR-parallel}, where we specify the photoelectron energy in each sideband for both hydrogen and helium targets. For other initial states, bound-continuum contributions may have a non-negligible role in defining the phase jump emission angles because intermediate states will have more than one possible angular momentum value.

\subsection{The non-parallel case, $\Theta\neq 0^{\circ}$}

\begin{figure*}
 \includegraphics[width=0.32\textwidth,clip]{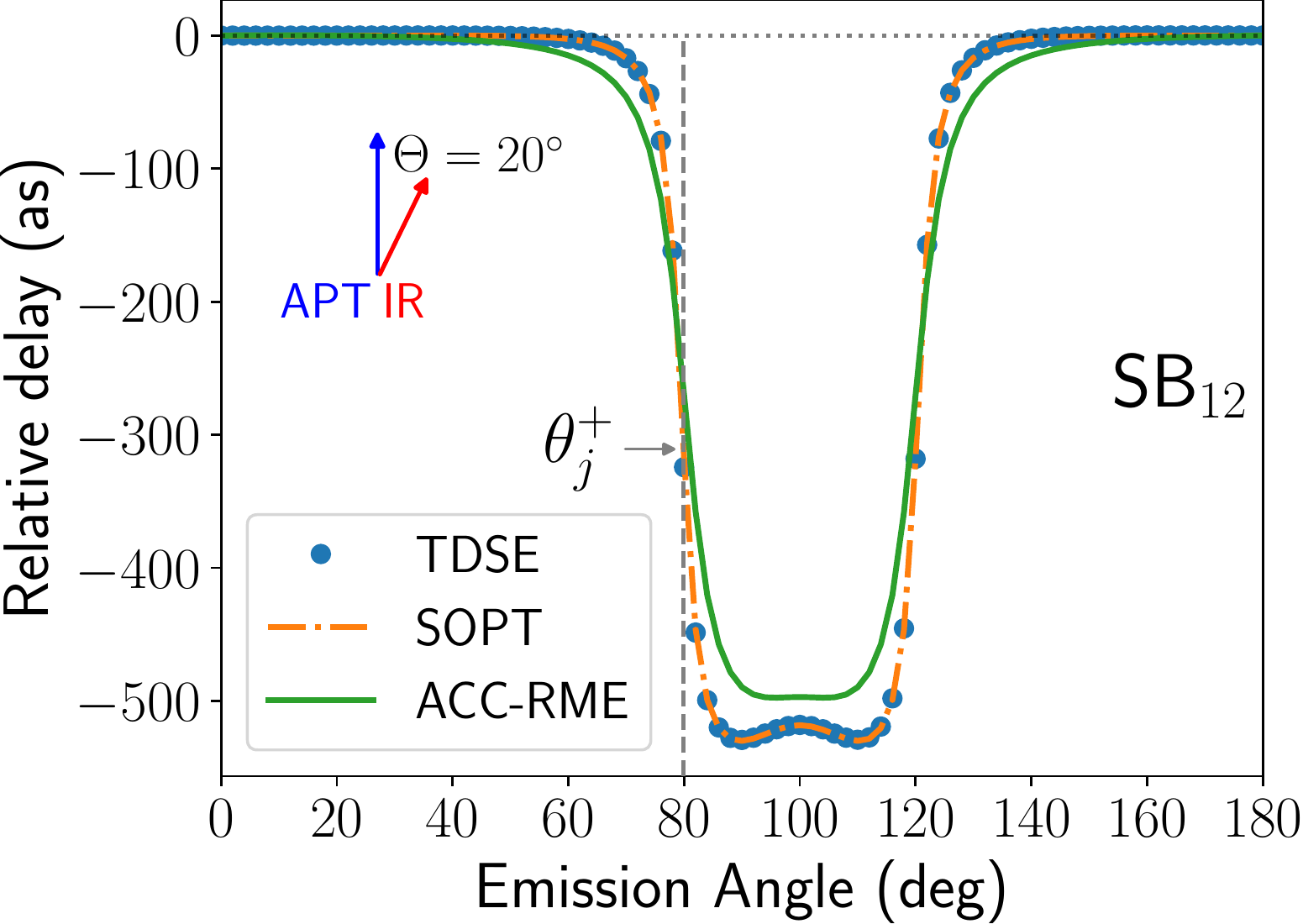}
 \includegraphics[width=0.32\textwidth,clip]{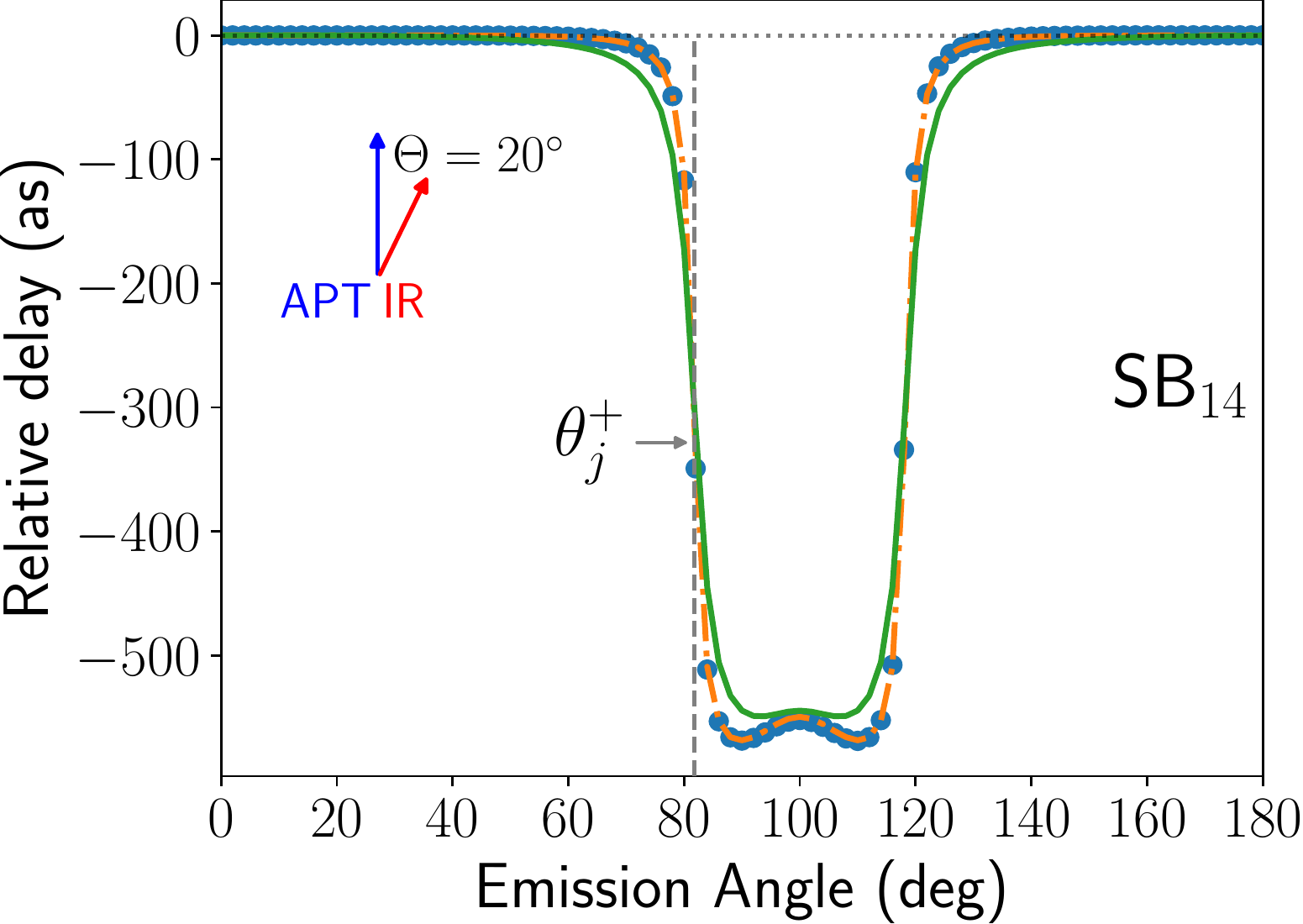}
 \includegraphics[width=0.32\textwidth,clip]{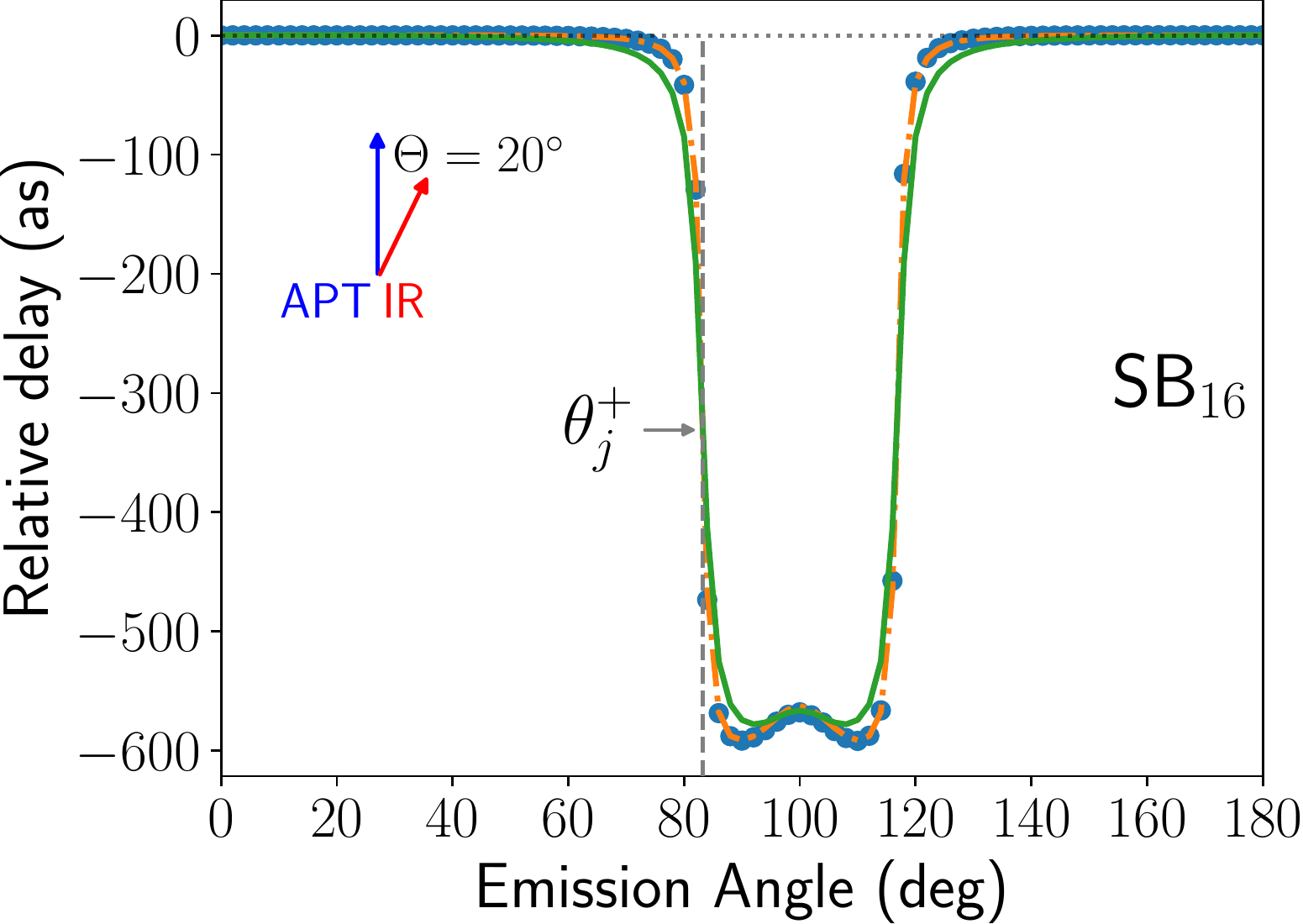}\\           
 \includegraphics[width=0.32\textwidth,clip]{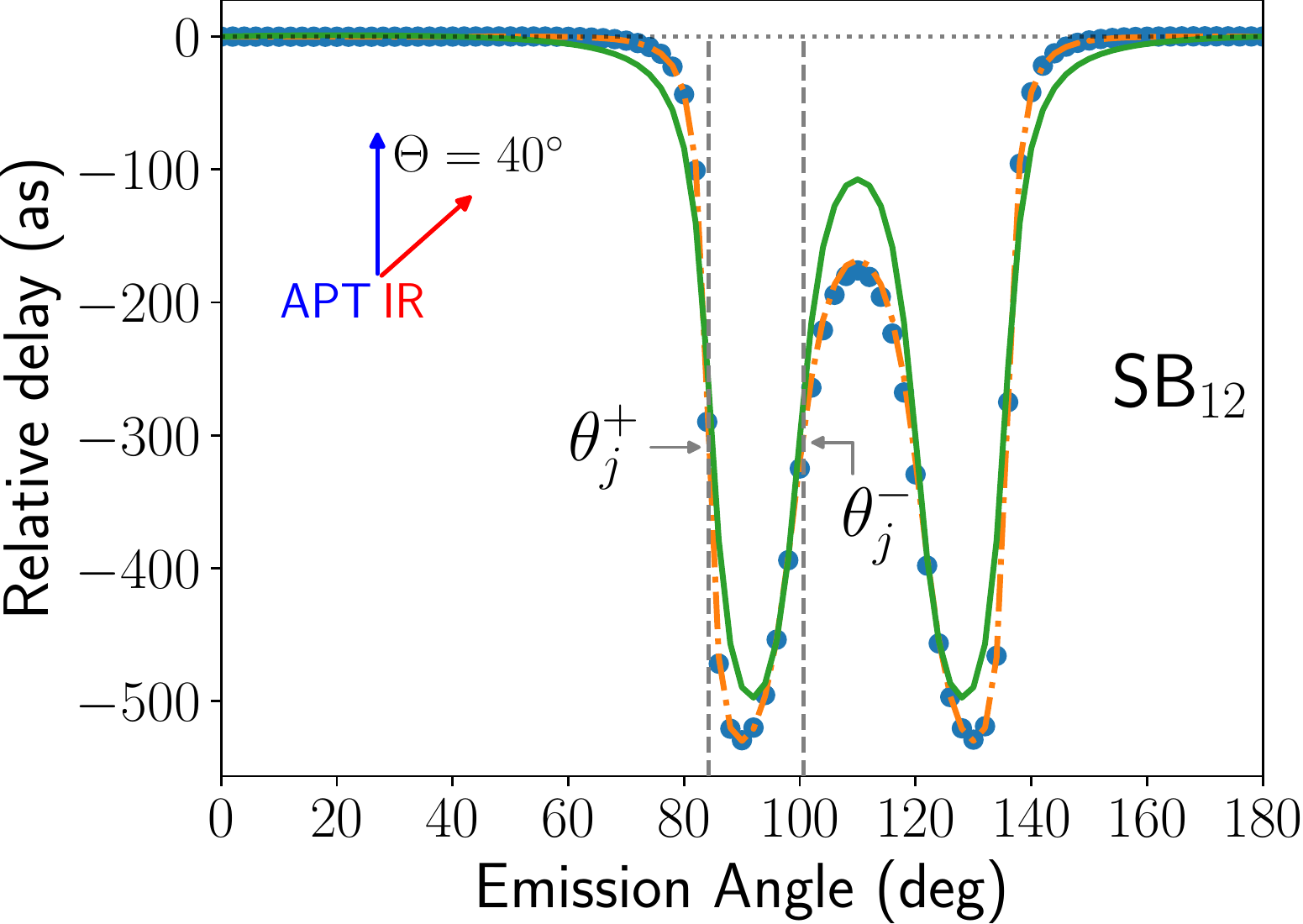}
 \includegraphics[width=0.32\textwidth,clip]{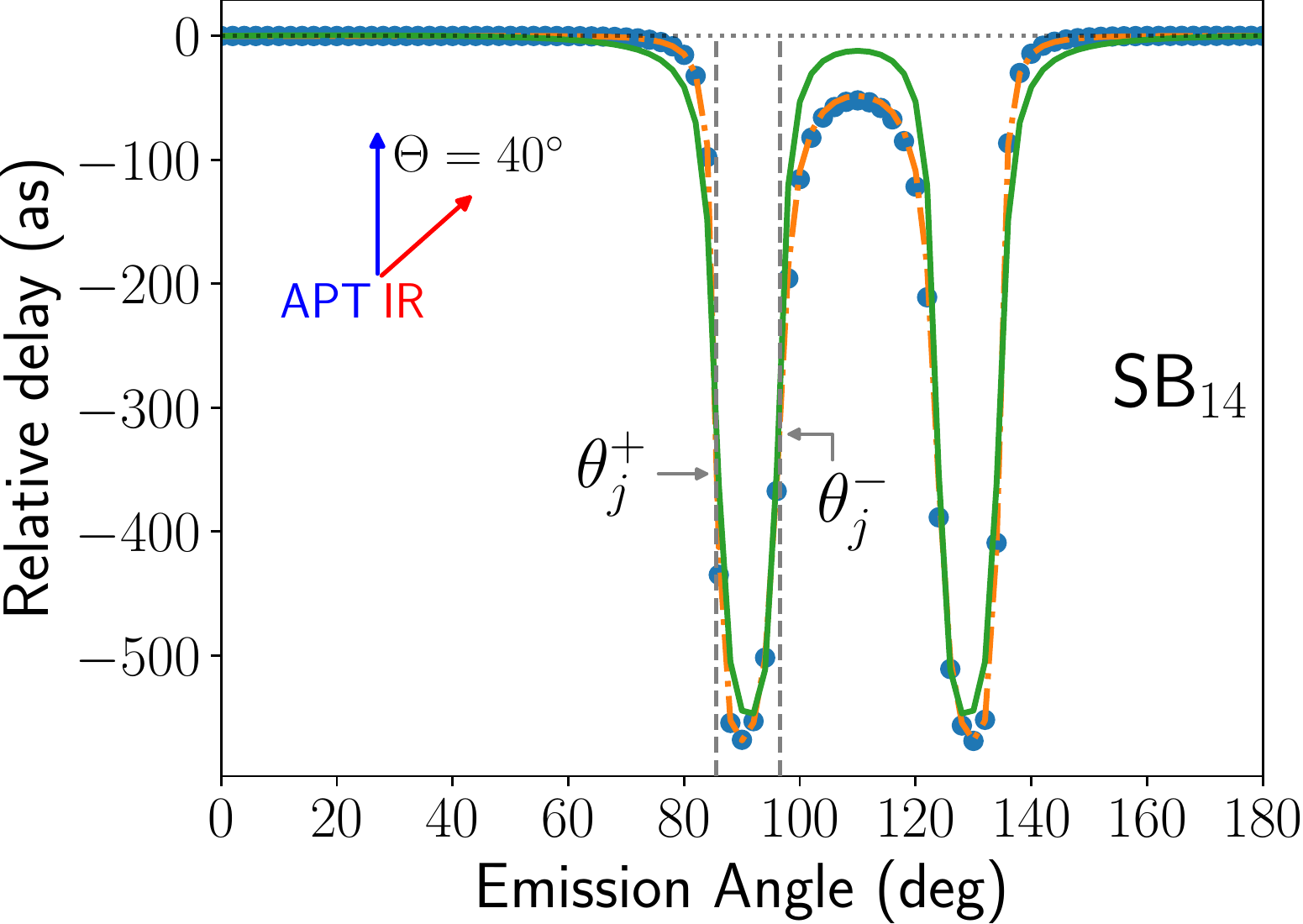}
 \includegraphics[width=0.32\textwidth,clip]{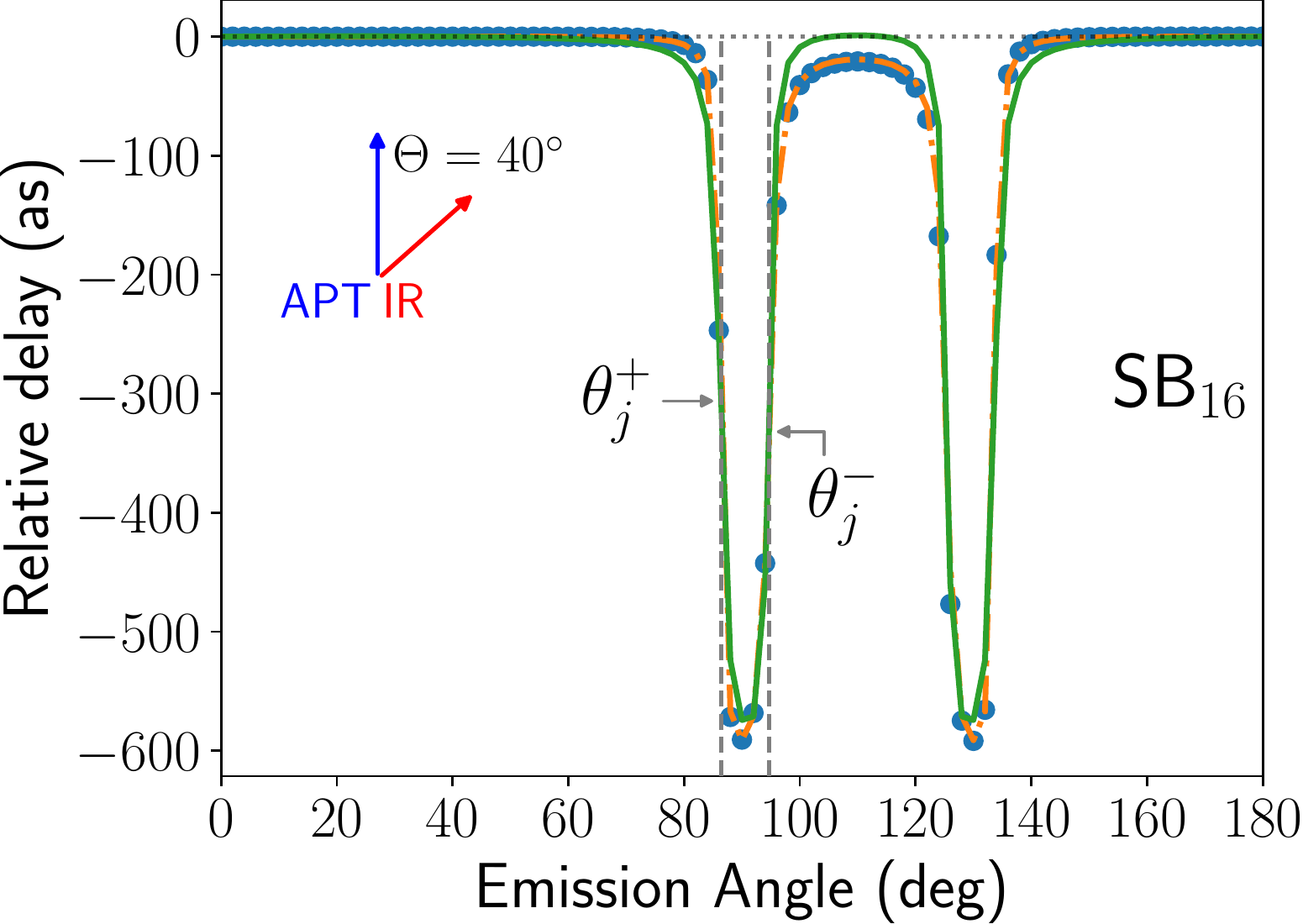}\\
 \includegraphics[width=0.32\textwidth,clip]{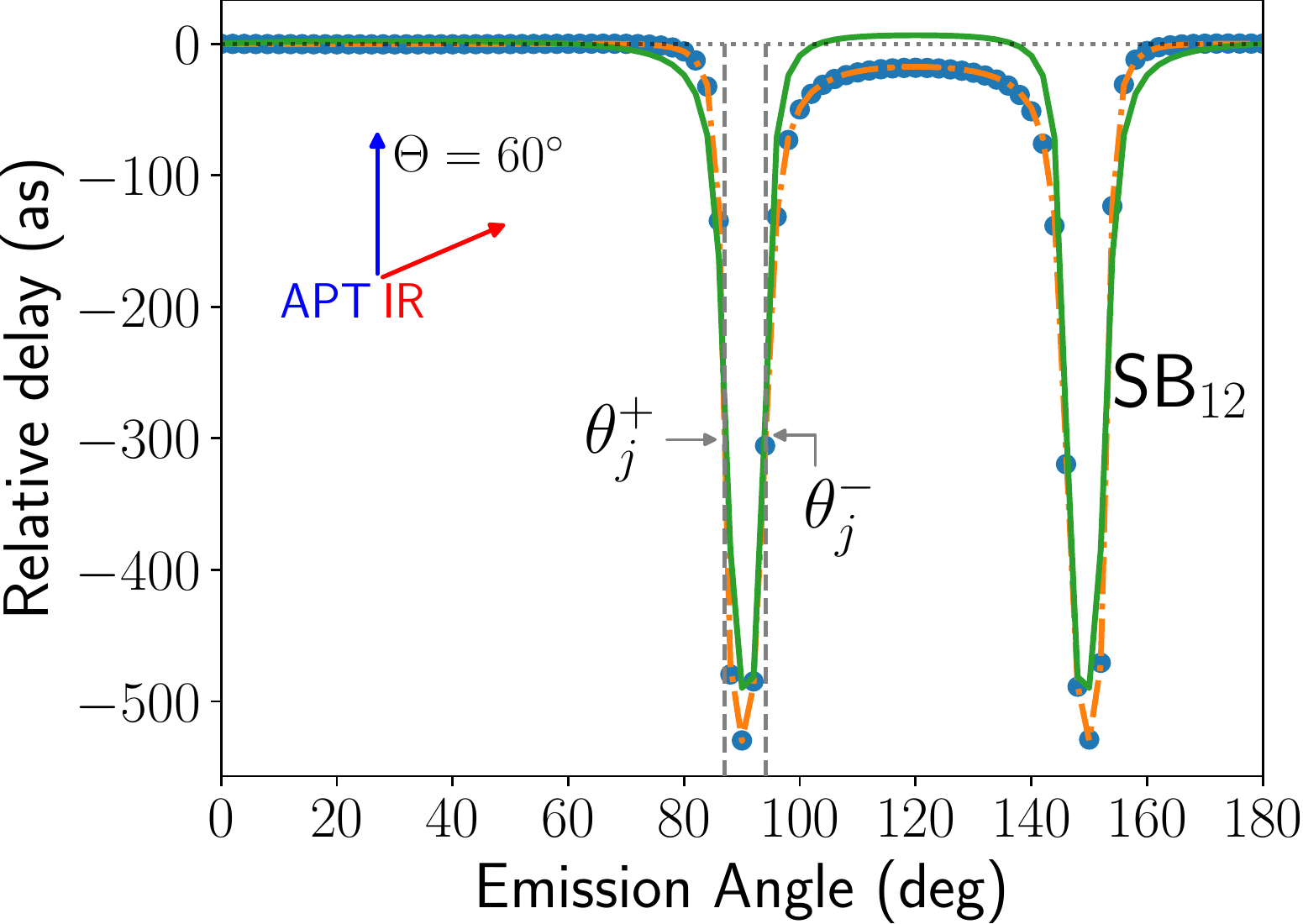}
 \includegraphics[width=0.32\textwidth,clip]{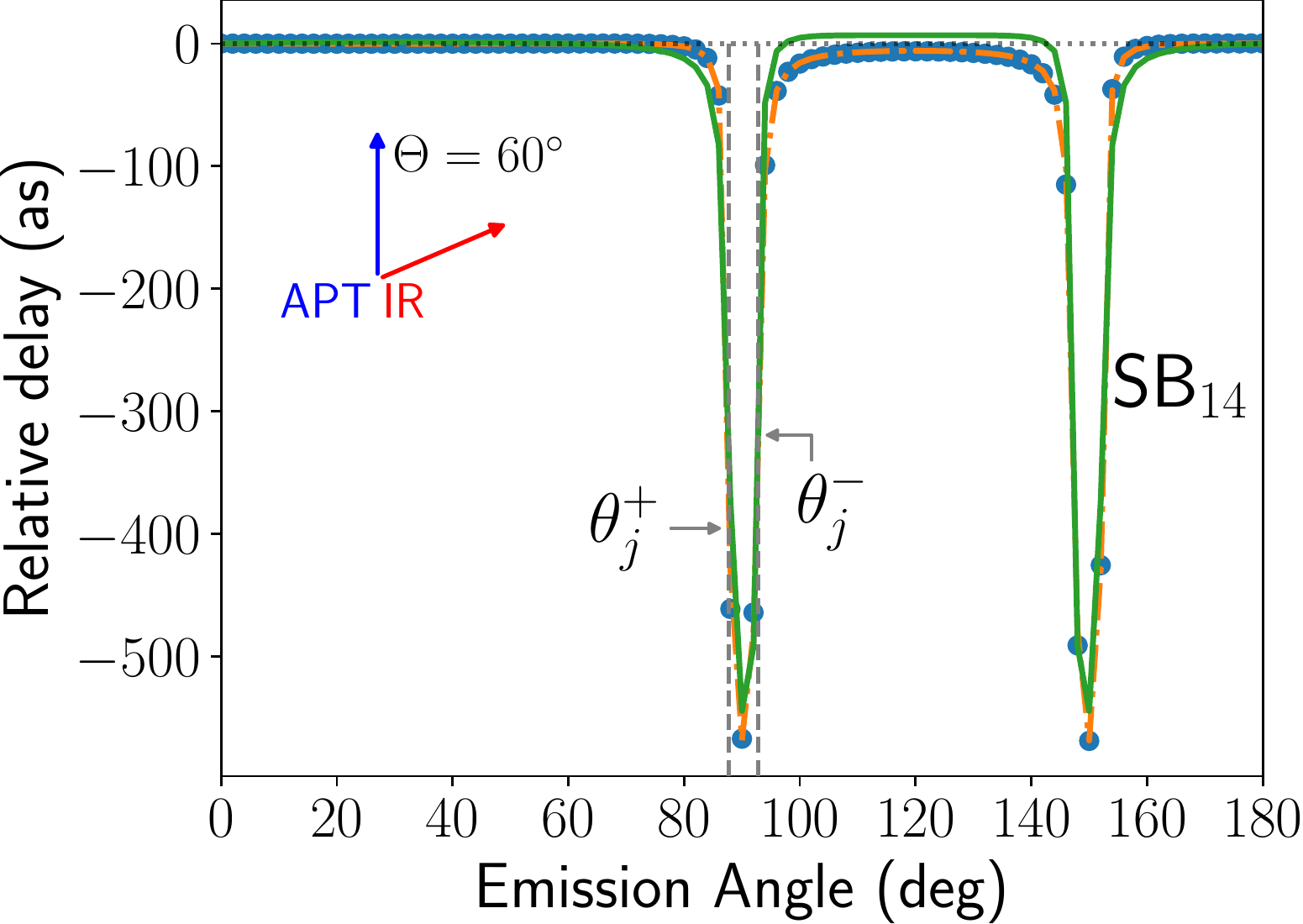}
 \includegraphics[width=0.32\textwidth,clip]{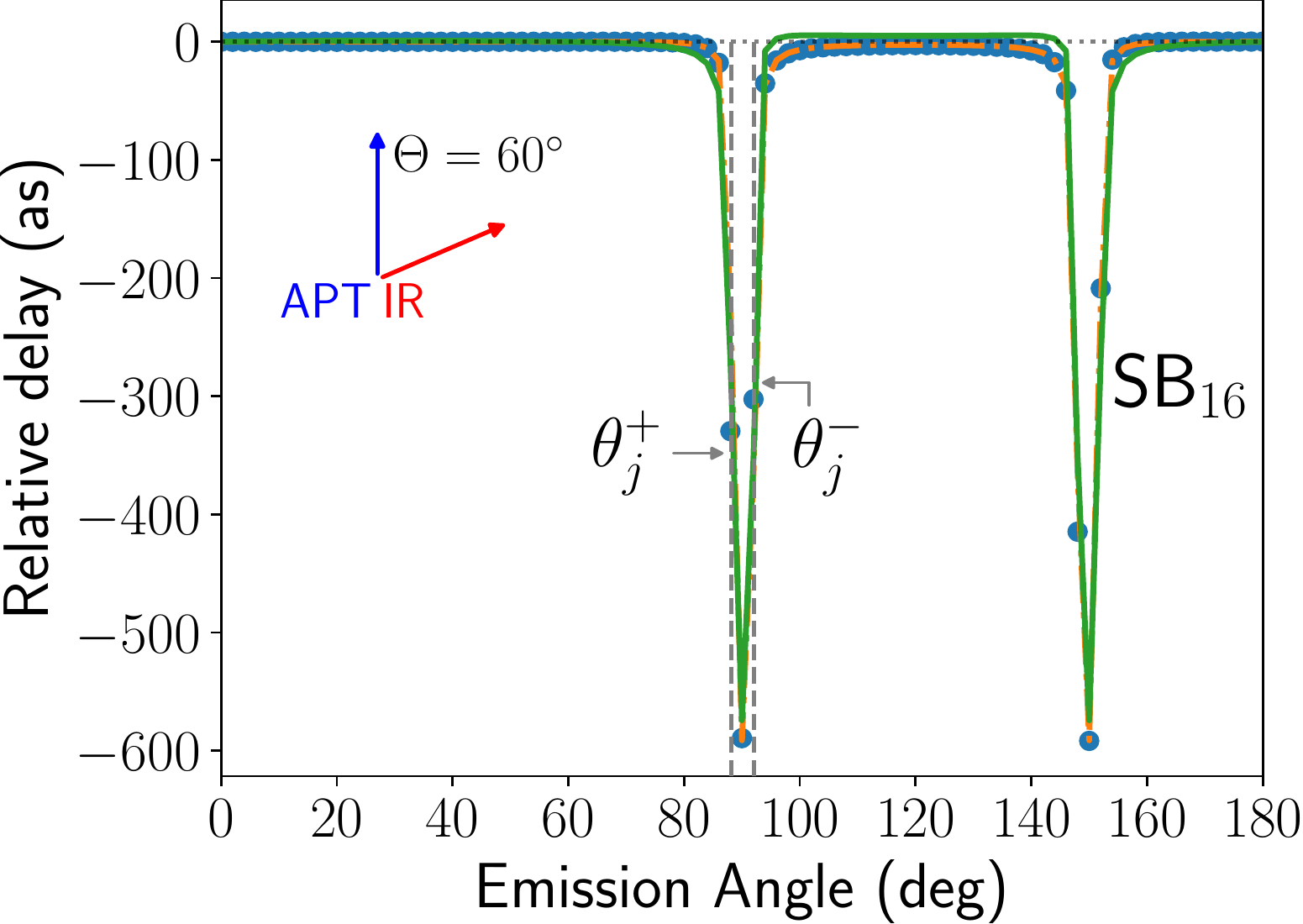}           
 \caption{Angularly-resolved atomic time delays at photoelectron energies coinciding with RABBITT sidebands for hydrogen atoms and different relative polarization angles $\Theta$. The results are computed through three different methods: TDSE, SOPT and ACC-RME calculations. The phase jump angles $\theta_j^{\pm}$ are obtained from Eq. \eqref{eq:angle-phase-jump-general}, and the relative time delay for that emission angle is obtained from linear interpolation of the TDSE results. This procedure may lead to larger fluctuations for sidebands with larger index due to the steeper variation of $\Delta \tau_{tot}$. We only display the first occurrence of the phase jump angles $\theta_j^{\pm}$ associated to absorption and emission pathways. The second phase jump angle for each family arises at emission angles symmetric with respect to $(\pi+\Theta)/2$.  \label{fig:AR-non-parallel}}
\end{figure*}

In the following, we explore the angularly resolved phase differences (equivalent, in terms of time delays) resulting from RABBITT spectra for different relative polarization angles $\Theta$ between attosecond pulse train and the IR field. In such a scheme, the light-atom system no longer has cylindrical symmetry, and atomic time delays depend on both the azimuthal and polar emission angles. We focus on electron emission into the plane defined by the polarization vectors of the XUV and IR fields. From now on, $\theta$ denotes the photoelectron emission angle in the $xy$-plane measured from the positive $x$-axis, the polarization direction of the attosecond pulse train (see Fig. \ref{fig:scheme}). 

In Fig. \ref{fig:AR-non-parallel}, we depict the angular variation of atomic time delays for sidebands 12 to 16 in hydrogen and different relative polarization angles $\Theta$ between the XUV and IR fields. We notice that a double-peak pattern unfolds as $\Theta$ increases.  The broad symmetric valley shapes found for collinear fields (Fig. \ref{fig:AR-parallel}), become two narrow shapes at specific angles (Fig. \ref{fig:AR-non-parallel}). To unveil the physical origin of this phenomenon, we generalize Eq. \eqref{eq:angle-phase-jump} to the case of non-collinear laser polarization directions. We find that phase jump angles satisfy the relation
\begin{align}\label{eq:angle-phase-jump-general}
\cos(2\theta_j^{\pm}-\Theta)\simeq - \left[ \frac{1}{3} +\frac{2}{3}\frac{\vert T_0^{\pm}\vert}{\vert T_2^{\pm}\vert}\right] \cos \Theta, 
\end{align} after setting $\Delta=0$. A distinctive feature of this equation, and perhaps the central contribution of this paper, is that it predicts the emergence of another family of phase jump angles. Briefly, the $\cos \Theta$ factor can make the right-hand side greater than minus one, regardless of the value for $\vert T_0^{\pm}/T_2^{\pm}\vert$. Therefore, it will be possible to find real solutions also for phase jump angles $\theta_j^{-}$, associated with emission pathways and forbidden in the parallel case. The availability of a second family of phase jump angles should, therefore, restore atomic time delays to values similar to those found before the first phase jump induced by the absorption pathways. Hence, changing the relative polarization angle can split the broad time delay valleys obtained for the parallel case into two narrower ones. For sidebands 12 to 16, we find that a relative polarization angle $\Theta$ larger than $\sim 35^{\circ}$ leads to physical solutions also for emission pathways, in agreement with the results in Fig. \ref{fig:AR-non-parallel}. Thus, the analytic expression in Eq. \eqref{eq:angle-phase-jump-general} predicts, in a simplified manner, all the phase jump angles that ultimately shape the angular evolution of atomic time delays.  
 
As in the parallel case, the vertical dashed lines mark the phase jump angles we calculate from Eq. \eqref{eq:angle-phase-jump-general} using SOPT radial matrix elements. Although we do not explicitly indicate them,  each family presents a second phase jump at emission angles symmetric with respect to $(\pi+\Theta)/2$. A plain visual inspection of Fig. \ref{fig:AR-non-parallel} leads us to conclude that solutions $\theta_j^{\pm}$ provide a reasonable approximation for the full width at half minima of the time delay valleys. Therefore, by using Eq. \eqref{eq:angle-phase-jump-general}, we can analyze the narrowing of the time delay valleys with increasing photoelectron energy or the relative polarization angle. For fixed $\Theta$ values, the quotient $\vert T_0^+/T_2^+ \vert$ rises for increasing photoelectron energy, as the inset in Fig. \ref{fig:AR-parallel} shows. In that case, the right-hand side of Eq. \eqref{eq:angle-phase-jump-general} will tend to $-\cos \Theta$ from above. In the asymptotic region, where $\vert T_0^+\vert \simeq \vert T_2^+\vert$, the first occurrence of the phase jump angle $\theta_j^+$ will approach $90^{\circ}$ from the left, \emph{i.e.} the phase jump will occur for electron emission perpendicular to the polarization vector of the attosecond pulse train. A similar analysis shows that the other phase jump enclosing the time delay valley will approach $90^{\circ}$ from the right. 

The previous analysis shows the role that radial degrees of freedom play in shaping the angular variation of atomic time delays through the radial matrix elements. Alternatively, the angular degrees of freedom may induce similar effects. Larger tilt angles $\Theta$ may also cause a narrowing of the time delay valleys we obtain for sidebands, as Fig. \ref{fig:AR-non-parallel} shows. On the assumption that photoelectron kinetic energy remains constant, larger $\Theta$ values make the right-hand side of Eq. \eqref{eq:angle-phase-jump-general} to approach zero from below because of the $\cos \Theta$ factor. At first glance, this factor should push the phase jump angle $\theta_j^+$ to the left. However, this effect is over-compensated by the $\Theta/2$ shift arising from the left-hand side of Eq. \eqref{eq:angle-phase-jump-general}. In sum, increasing the tilt angle $\Theta$ has a net effect on the width of time delay valleys similar to considering higher photoelectron kinetic energies. Consequently, this may explain the unexpected accuracy improvement of the ACC-RME results for larger $\Theta$ values. 

In agreement with recent experimental findings \cite{Jiang2022}, our results in Fig. \ref{fig:AR-non-parallel} show that total time delay minima arise for electron emission perpendicular to either XUV or IR polarization vectors, irrespective of the tilt angle $\Theta$ and photoelectron kinetic energy. However, low photoelectron emission probabilities usually hinder the direct observation of such minima \cite{Heuser2016,Jiang2022}. A more efficient approach to experimentally verify their existence is to induce, on the system, the double phase-jumps we unveil. Contrary to what Fig. \ref{fig:AR-non-parallel} might suggest, the emergence of the double-well structure is not restricted to tilt angles larger than $35^{\circ}$. The interplay of radial and angular degrees of freedom evidenced in Eq. \eqref{eq:angle-phase-jump-general} make it possible to find that signature for every angle $\Theta\neq 0$, provided that photoelectron kinetic energy $\varepsilon_{k_e}$ is high enough. Nonetheless, for asymptotically large $\varepsilon_{k_e}$ values, the quotient of radial matrix elements approaches unity with the consequent shrink in the width of the wells, making them harder (if not impossible) to be experimentally detected. Clearly, photoelectron emission probabilities play a central role in studying this phenomenon.
 
Next, we focus on how the relative polarization angle $\Theta$ modifies the angular emission pattern. With that goal in mind, we concentrate on the $A$ parameter in Eq. \eqref{eq:SB-signal-fit} to estimate how the tilt angle $\Theta$ affects photoemission probabilities in different setups. For infinitely long IR fields, the $A$ parameter is nothing but the zeroth-order Fourier series coefficient for the periodic dependence of the sideband signal on the delay $\tau$. Therefore, it is a robust (almost time-independent) measure of the differential photoelectron emission probabilities.

\begin{figure}
 \includegraphics[width=\columnwidth,clip]{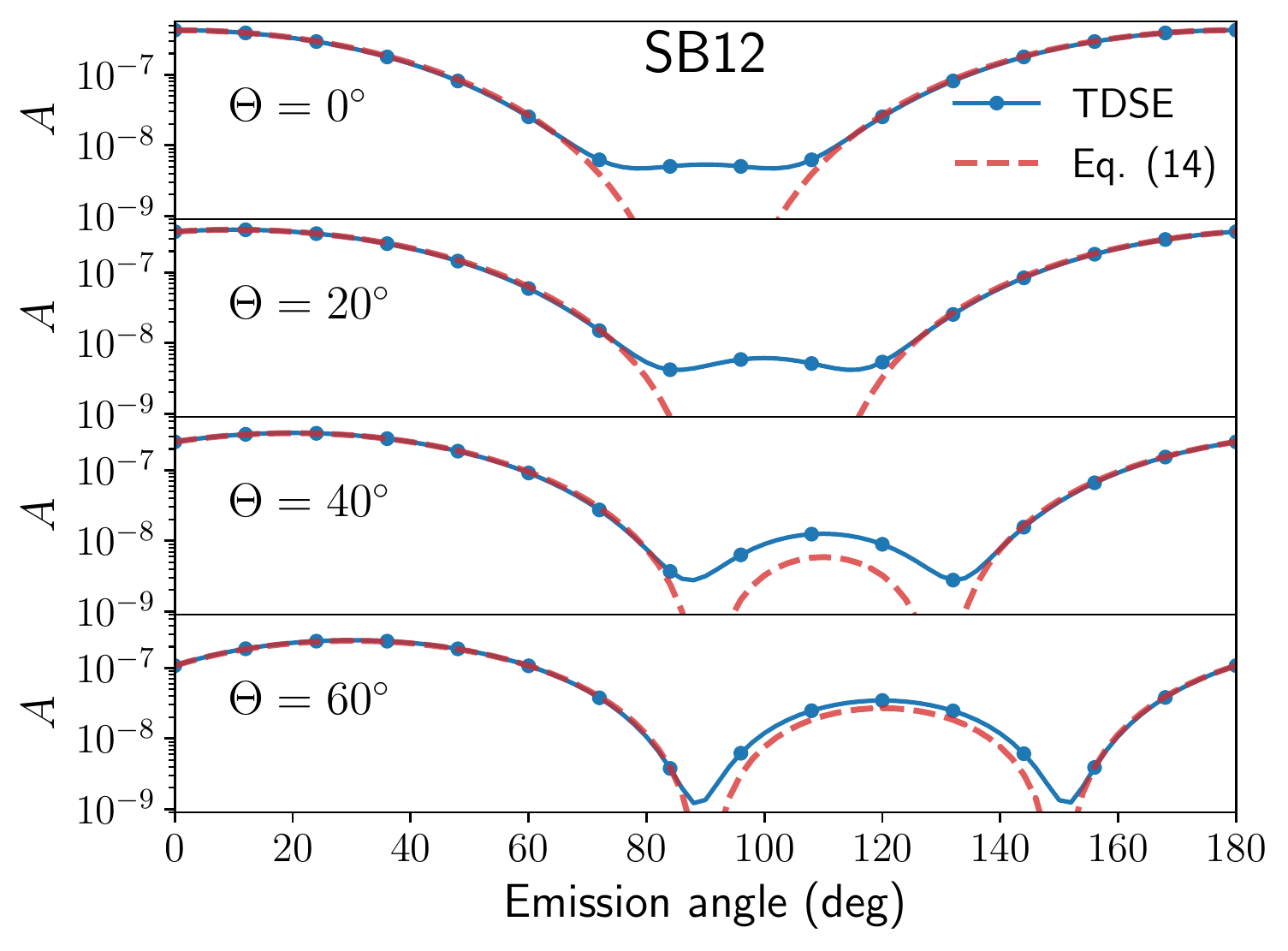}
 \caption{Time Delay independent photoelectron emission probabilities ($A$ term in Eq. \eqref{eq:SB-signal-fit}), as a function of photoelectron emission angle $\theta$ for different relative polarization angles $\Theta$ between the XUV and the IR fields. 
\label{fig:prob-analysis}}
\end{figure}

In fig. \ref{fig:prob-analysis}, we show the dependence of the $A$ parameter on the photoelectron emission angle $\theta$ for different pump-probe relative polarization angles. The solid line and circles represent the results we obtain by fitting angle-resolved photoelectron spectra from TDSE simulations with Eq. \eqref{eq:SB-signal-fit}. The data unambiguously show a signal decrease for electron emission in directions perpendicular to the polarization vectors of XUV or IR laser fields. The trend becomes apparent for increasing relative polarization angles $\Theta$, with deeper minima in the signal. The photoelectron emission probabilities span two orders of magnitude and explain the difficulties in the detection, particularly for emission directions where atomic time delay minima arise. 

To further understand this behavior, we again resort to analytic tools. Based on the soft-photon approximation \cite{Maquet2007}, it is possible to show that the angular dependence of the non-oscillatory factor $A$ in Eq. \eqref{eq:SB-signal-fit} is proportional to $\cos^4\theta$, in the parallel $(\Theta=0^{\circ})$ case \cite{Ivanov2017}. This result is formally valid for asymptotically high photoelectron energies. However, it describes the angular dependence of $A$ for $\varepsilon_{k_e}$ values above a few electronvolts \cite{Picard2014,Ivanov2017}. To our knowledge, a similar expression for the non-parallel case has not been reported in the literature. In the following, we derive it from previous results. In Ref. \cite{Boll2016,Boll2017}, we model the sideband photoelectron angular distributions for different RABBITT-like configurations as,

\begin{align}\label{eq:2-center-SB} 
I_{2q}(\theta,\phi) \propto [1+\beta P_2(\cos \theta)] \sin^{2}\left(\frac{\mathbf{k}_{e}\cdot \mathbf{R}}{2}\right), 
\end{align}with $\beta$ being the asymmetry parameter for the single-photon ionization of the atomic target by the XUV pulse (pump stage), $P_2(x)$  the second-order Legendre polynomial, $\mathbf{k}_{e}$ the photoelectron momentum, and $\mathbf{R}$ a vector quantity collinear to the IR polarization direction (See Ref. \cite{Boll2016,Boll2017} for further details). At variance with more standard approaches like Eq. \eqref{eq:SB-signal-fit}, the equation above remains valid for IR intensities such that cc transitions beyond first-order significantly contribute to sideband signals. For usual RABBITT setups, with only lowest order cc channels effectively open, the relation $\vert \mathbf{k}_{e}\cdot \mathbf{R}\vert \ll 1$ holds, and the rightmost factor in Eq. \eqref{eq:2-center-SB} can be approximated by $\sin^2(x)\sim x^2$. Besides, as we are analyzing electron emission in the plane defined by the XUV and IR polarization vectors, the scalar product $\mathbf{k}_{e}\cdot \mathbf{R}$ reduces to $k_{e}R\, (\cos\theta \cos\Theta+\sin\theta\sin\Theta) = k_{e}R\cos(\theta-\Theta)$. With these results in mind, and provided that $\beta=2$ for initial $s$ states, it is straightforward to show that the angular dependence of the $A$ parameter in Eq. \eqref{eq:SB-signal-fit} reads 
\begin{align}\label{eq:A-full}
A\propto \cos^2\theta \cos^2 (\theta-\Theta),
\end{align} for every $\Theta$ value, and it recovers the $\cos^4\theta$ law for a parallel configuration of the fields. 

The direct comparison between the asymptotic model in Eq. \eqref{eq:A-full} (red dashed line) and TDSE simulations in Fig. \ref{fig:prob-analysis} highlights several facets of the process. First, the angular algebra largely dictates the salient features of photoelectron emission probability. Thus, we expect a universal character for the $A$ parameter, provided that the initial atomic state has the same angular quantum numbers. Second, the enhanced predictive power of Eq. \eqref{eq:A-full} for $\Theta$ approaching $90^{\circ}$, brings to the fore the fact that only $d$ partial waves survive for such a configuration \cite{Meyer2008,Boll2020}. Therefore, the $s$ partial wave depletion makes the unbalance of radial matrix elements to play a diminishing role. Such an arrangement of laser fields represents the best opportunities for prediction models based on the soft-photon approximation, as they cannot account for relative variations in the weight of each partial wave. In turn, deviations in Fig. \ref{fig:prob-analysis} between TDSE results and those from Eq. \eqref{eq:A-full} also stem from small phase shifts between partial waves. However, model and accurate calculations differ for photoelectron emission directions whose probabilities are too low for its detection with current experimental capabilities. This fact renders soft-photon approximation models an ideal candidate for quickly assessing some gross features in angularly-resolved two-color reaction processes.        
       
\section{Conclusions and outlook}\label{sec:conclusions}
A simple analytical model is presented in order to predict angularly resolved photoionization time delays in atomic targets employing the RABBITT technique with different relative polarization angles between the XUV and the IR fields. The model reproduces the significant variation of the time delays with the photoelectron angle emission in H and He atoms, revealing a clear phase jump at around $65^{\circ}$, discussed in previous works \cite{Heuser2016,Ivanov2017,Busto2019}. This jump is associated to the relative contribution of different partial waves. Our analytical model perfectly reproduces this behaviour. It is demonstrated in hydrogenic atoms, and compared with previous theoretical and experimental data for a standard RABBITT scheme, \emph{i.e.}, collinear fields. More interestingly, we further analyze the results for different relative polarization angles. 

First, it is found that it is possible to manipulate the dependencies of the relative time delays by varying the polarization angle between the fields. Indeed, a characteristic double-well shape is found for the time delays as a function of this angle. In other words, a new set of phase jump angles is found for non-collinear fields. We explicitly show that radial matrix elements of emission pathways dictate the photoelectron emission angles for which this new family of phase jumps occurs. Furthermore, we demonstrate that phase jumps associated with emission pathways are forbidden only for parallel orientation of the XUV and IR polarization vectors. For any other relative polarization angle between the XUV and IR fields, it should be possible to observe the double-well structure. However, for sufficiently high photoelectron kinetic energy, its observation may be hindered by the shrinking of the characteristic width of each well which, in turn, can be predicted from the radial matrix elements (quotient) dependence on kinetic energy.    

Then, we study the angular dependence of the fitting parameter $A$  characterizing photoelectron emission probabilities in RABBITT sidebands. Taking advantage of previous studies, we put forward an asymptotic law able to predict the behavior of this fitting parameter with reasonably good accuracy. In turn, this simple asymptotic law shows the deep connection between time delay minima for electron emission directions perpendicular to either the XUV or the IR laser fields and the angular degrees of freedom of the system.

Finally, we want to point out that our analytic model admits an extension to study angular variations of time delays on more complex atomic (or even molecular) targets. That goal can be achieved provided that single-center expansions for bound-continuum dipole matrix elements are available. We expect this study may foster renewed experimental and theoretical efforts to understand the information conveyed by these observables and, eventually, use that knowledge to control their behavior.            

\begin{acknowledgments}
D.B., L.M, and O.F. acknowledge financial support from the Agencia Nacional de Promoción Científica y Tecnológica, Project PICT No. 01912 (2015-3392), and from the Consejo Nacional de Investigaciones Científicas y Técnicas de la República Argentina, Project PIP No. 3245 (11220200103245). A.P. acknowledges the funding from the PRICIT Program from Comunidad de Madrid (Ref. PCD-I3PCD026 UAM), and Project No. FIS2017-92382-EXP from the Ministerio de Economía y Competitividad (Spain). Part of the results presented in this paper have been obtained by using the facilities of the CCT-Rosario Computational Center, member of the High Performance Computing National System (SNCAD, MincyT-Argentina).
\end{acknowledgments}

\appendix*
\section{Expansion coefficients $g_{L,L'}$}\label{sec:appendix}
In this appendix we explicitely write the coefficients $g_{L,L'}$ for the double summation in Eq. \eqref{eq:oscillatory-terms} defining the oscillatory pattern of sidebands. The same set of factors enter Eq. \eqref{eq:trigo-phase} from which we extract the time delay $\tau_{at}$. As we solve the TDSE with the \emph{Qprop} code, and it allows to set non-collinear fields only if their polarization vectors lie in the $xy$ plane, these results were obtained for that  configuration of the fields. Besides, as we consider emission into the plane $xy$ defined by the polarization vectors of the XUV and IR radiation fields, the following results are specialized for that case.
\begin{align}
g_{0,0}=&\frac{1}{2}\cos^2\Theta\\
g_{0,2}=&\frac{1}{4}\cos\Theta\left[\cos\Theta+3\cos(\Theta-2\theta)\right]\\
g_{2,0}=&g_{0,2}\\
\begin{split}
g_{2,2}=&\frac{1}{8}\cos\Theta\left[\cos\Theta+6\cos(\Theta-2\theta)\right]\\&+\frac{9}{16}\left[1+\cos(2\Theta-4\theta)\right]
\end{split}
\end{align}  
Using these expressions for $g_{L,L'}$ it is possible to obtain the angularly resolved time delays from Eq. \eqref{eq:trigo-phase}, by replacing the radial matrix elements obtained from SOPT or ACC-RME calculations.

\bibliography{cc-phases}

\begin{thebibliography}{43}%
\makeatletter
\providecommand \@ifxundefined [1]{%
 \@ifx{#1\undefined}
}%
\providecommand \@ifnum [1]{%
 \ifnum #1\expandafter \@firstoftwo
 \else \expandafter \@secondoftwo
 \fi
}%
\providecommand \@ifx [1]{%
 \ifx #1\expandafter \@firstoftwo
 \else \expandafter \@secondoftwo
 \fi
}%
\providecommand \natexlab [1]{#1}%
\providecommand \enquote  [1]{``#1''}%
\providecommand \bibnamefont  [1]{#1}%
\providecommand \bibfnamefont [1]{#1}%
\providecommand \citenamefont [1]{#1}%
\providecommand \href@noop [0]{\@secondoftwo}%
\providecommand \href [0]{\begingroup \@sanitize@url \@href}%
\providecommand \@href[1]{\@@startlink{#1}\@@href}%
\providecommand \@@href[1]{\endgroup#1\@@endlink}%
\providecommand \@sanitize@url [0]{\catcode `\\12\catcode `\$12\catcode
  `\&12\catcode `\#12\catcode `\^12\catcode `\_12\catcode `\%12\relax}%
\providecommand \@@startlink[1]{}%
\providecommand \@@endlink[0]{}%
\providecommand \url  [0]{\begingroup\@sanitize@url \@url }%
\providecommand \@url [1]{\endgroup\@href {#1}{\urlprefix }}%
\providecommand \urlprefix  [0]{URL }%
\providecommand \Eprint [0]{\href }%
\providecommand \doibase [0]{https://doi.org/}%
\providecommand \selectlanguage [0]{\@gobble}%
\providecommand \bibinfo  [0]{\@secondoftwo}%
\providecommand \bibfield  [0]{\@secondoftwo}%
\providecommand \translation [1]{[#1]}%
\providecommand \BibitemOpen [0]{}%
\providecommand \bibitemStop [0]{}%
\providecommand \bibitemNoStop [0]{.\EOS\space}%
\providecommand \EOS [0]{\spacefactor3000\relax}%
\providecommand \BibitemShut  [1]{\csname bibitem#1\endcsname}%
\let\auto@bib@innerbib\@empty
\bibitem [{\citenamefont {Pazourek}\ \emph {et~al.}(2015)\citenamefont
  {Pazourek}, \citenamefont {Nagele},\ and\ \citenamefont
  {Burgd\"orfer}}]{Pazourek2015}%
  \BibitemOpen
  \bibfield  {author} {\bibinfo {author} {\bibfnamefont {R.}~\bibnamefont
  {Pazourek}}, \bibinfo {author} {\bibfnamefont {S.}~\bibnamefont {Nagele}},\
  and\ \bibinfo {author} {\bibfnamefont {J.}~\bibnamefont {Burgd\"orfer}},\
  }\bibfield  {title} {\bibinfo {title} {Attosecond chronoscopy of
  photoemission},\ }\href {https://doi.org/10.1103/RevModPhys.87.765}
  {\bibfield  {journal} {\bibinfo  {journal} {Rev. Mod. Phys.}\ }\textbf
  {\bibinfo {volume} {87}},\ \bibinfo {pages} {765} (\bibinfo {year}
  {2015})}\BibitemShut {NoStop}%
\bibitem [{\citenamefont {Hentschel}\ \emph {et~al.}(2001)\citenamefont
  {Hentschel}, \citenamefont {Kienberger}, \citenamefont {Spielmann},
  \citenamefont {Reider}, \citenamefont {Milosevic}, \citenamefont {Brabec},
  \citenamefont {Corkum}, \citenamefont {Heinzmann}, \citenamefont {Drescher},\
  and\ \citenamefont {Krausz}}]{Hentschel2001}%
  \BibitemOpen
  \bibfield  {author} {\bibinfo {author} {\bibfnamefont {M.}~\bibnamefont
  {Hentschel}}, \bibinfo {author} {\bibfnamefont {R.}~\bibnamefont
  {Kienberger}}, \bibinfo {author} {\bibfnamefont {C.}~\bibnamefont
  {Spielmann}}, \bibinfo {author} {\bibfnamefont {G.~A.}\ \bibnamefont
  {Reider}}, \bibinfo {author} {\bibfnamefont {N.}~\bibnamefont {Milosevic}},
  \bibinfo {author} {\bibfnamefont {T.}~\bibnamefont {Brabec}}, \bibinfo
  {author} {\bibfnamefont {P.}~\bibnamefont {Corkum}}, \bibinfo {author}
  {\bibfnamefont {U.}~\bibnamefont {Heinzmann}}, \bibinfo {author}
  {\bibfnamefont {M.}~\bibnamefont {Drescher}},\ and\ \bibinfo {author}
  {\bibfnamefont {F.}~\bibnamefont {Krausz}},\ }\bibfield  {title} {\bibinfo
  {title} {Attosecond metrology},\ }\href {https://doi.org/10.1038/35107000}
  {\bibfield  {journal} {\bibinfo  {journal} {Nature}\ }\textbf {\bibinfo
  {volume} {414}},\ \bibinfo {pages} {509} (\bibinfo {year}
  {2001})}\BibitemShut {NoStop}%
\bibitem [{\citenamefont {V\'eniard}\ \emph {et~al.}(1996)\citenamefont
  {V\'eniard}, \citenamefont {Ta\"{\i}eb},\ and\ \citenamefont
  {Maquet}}]{Veniard1996}%
  \BibitemOpen
  \bibfield  {author} {\bibinfo {author} {\bibfnamefont {V.}~\bibnamefont
  {V\'eniard}}, \bibinfo {author} {\bibfnamefont {R.}~\bibnamefont
  {Ta\"{\i}eb}},\ and\ \bibinfo {author} {\bibfnamefont {A.}~\bibnamefont
  {Maquet}},\ }\bibfield  {title} {\bibinfo {title} {Phase dependence of
  (n+1)-color (n>1) ir-xuv photoionization of atoms with higher harmonics},\
  }\href {https://doi.org/10.1103/PhysRevA.54.721} {\bibfield  {journal}
  {\bibinfo  {journal} {Phys. Rev. A}\ }\textbf {\bibinfo {volume} {54}},\
  \bibinfo {pages} {721} (\bibinfo {year} {1996})}\BibitemShut {NoStop}%
\bibitem [{\citenamefont {Paul}\ \emph {et~al.}(2001)\citenamefont {Paul},
  \citenamefont {Toma}, \citenamefont {Breger}, \citenamefont {Mullot},
  \citenamefont {Augé}, \citenamefont {Balcou}, \citenamefont {Muller},\ and\
  \citenamefont {Agostini}}]{Paul2001}%
  \BibitemOpen
  \bibfield  {author} {\bibinfo {author} {\bibfnamefont {P.~M.}\ \bibnamefont
  {Paul}}, \bibinfo {author} {\bibfnamefont {E.~S.}\ \bibnamefont {Toma}},
  \bibinfo {author} {\bibfnamefont {P.}~\bibnamefont {Breger}}, \bibinfo
  {author} {\bibfnamefont {G.}~\bibnamefont {Mullot}}, \bibinfo {author}
  {\bibfnamefont {F.}~\bibnamefont {Augé}}, \bibinfo {author} {\bibfnamefont
  {P.}~\bibnamefont {Balcou}}, \bibinfo {author} {\bibfnamefont {H.~G.}\
  \bibnamefont {Muller}},\ and\ \bibinfo {author} {\bibfnamefont
  {P.}~\bibnamefont {Agostini}},\ }\bibfield  {title} {\bibinfo {title}
  {Observation of a train of attosecond pulses from high harmonic generation},\
  }\href {https://doi.org/10.1126/science.1059413} {\bibfield  {journal}
  {\bibinfo  {journal} {Science}\ }\textbf {\bibinfo {volume} {292}},\ \bibinfo
  {pages} {1689} (\bibinfo {year} {2001})}\BibitemShut {NoStop}%
\bibitem [{\citenamefont {{de Carvalho}}\ and\ \citenamefont
  {Nussenzveig}(2002)}]{deCarvalho2002}%
  \BibitemOpen
  \bibfield  {author} {\bibinfo {author} {\bibfnamefont {C.}~\bibnamefont {{de
  Carvalho}}}\ and\ \bibinfo {author} {\bibfnamefont {H.}~\bibnamefont
  {Nussenzveig}},\ }\bibfield  {title} {\bibinfo {title} {Time delay},\ }\href
  {https://doi.org/https://doi.org/10.1016/S0370-1573(01)00092-8} {\bibfield
  {journal} {\bibinfo  {journal} {Physics Reports}\ }\textbf {\bibinfo {volume}
  {364}},\ \bibinfo {pages} {83} (\bibinfo {year} {2002})}\BibitemShut
  {NoStop}%
\bibitem [{\citenamefont {Cavalieri}\ \emph {et~al.}(2007)\citenamefont
  {Cavalieri}, \citenamefont {M{\"u}ller}, \citenamefont {Uphues},
  \citenamefont {Yakovlev}, \citenamefont {Baltu{\v{s}}ka}, \citenamefont
  {Horvath}, \citenamefont {Schmidt}, \citenamefont {Bl{\"u}mel}, \citenamefont
  {Holzwarth}, \citenamefont {Hendel}, \citenamefont {Drescher}, \citenamefont
  {Kleineberg}, \citenamefont {Echenique}, \citenamefont {Kienberger},
  \citenamefont {Krausz},\ and\ \citenamefont {Heinzmann}}]{Cavalieri2007}%
  \BibitemOpen
  \bibfield  {author} {\bibinfo {author} {\bibfnamefont {A.~L.}\ \bibnamefont
  {Cavalieri}}, \bibinfo {author} {\bibfnamefont {N.}~\bibnamefont
  {M{\"u}ller}}, \bibinfo {author} {\bibfnamefont {T.}~\bibnamefont {Uphues}},
  \bibinfo {author} {\bibfnamefont {V.~S.}\ \bibnamefont {Yakovlev}}, \bibinfo
  {author} {\bibfnamefont {A.}~\bibnamefont {Baltu{\v{s}}ka}}, \bibinfo
  {author} {\bibfnamefont {B.}~\bibnamefont {Horvath}}, \bibinfo {author}
  {\bibfnamefont {B.}~\bibnamefont {Schmidt}}, \bibinfo {author} {\bibfnamefont
  {L.}~\bibnamefont {Bl{\"u}mel}}, \bibinfo {author} {\bibfnamefont
  {R.}~\bibnamefont {Holzwarth}}, \bibinfo {author} {\bibfnamefont
  {S.}~\bibnamefont {Hendel}}, \bibinfo {author} {\bibfnamefont
  {M.}~\bibnamefont {Drescher}}, \bibinfo {author} {\bibfnamefont
  {U.}~\bibnamefont {Kleineberg}}, \bibinfo {author} {\bibfnamefont {P.~M.}\
  \bibnamefont {Echenique}}, \bibinfo {author} {\bibfnamefont {R.}~\bibnamefont
  {Kienberger}}, \bibinfo {author} {\bibfnamefont {F.}~\bibnamefont {Krausz}},\
  and\ \bibinfo {author} {\bibfnamefont {U.}~\bibnamefont {Heinzmann}},\
  }\bibfield  {title} {\bibinfo {title} {Attosecond spectroscopy in condensed
  matter},\ }\href {https://doi.org/10.1038/nature06229} {\bibfield  {journal}
  {\bibinfo  {journal} {Nature}\ }\textbf {\bibinfo {volume} {449}},\ \bibinfo
  {pages} {1029} (\bibinfo {year} {2007})}\BibitemShut {NoStop}%
\bibitem [{\citenamefont {Schultze}\ \emph {et~al.}(2010)\citenamefont
  {Schultze}, \citenamefont {Fie{\ss}}, \citenamefont {Karpowicz},
  \citenamefont {Gagnon}, \citenamefont {Korbman}, \citenamefont {Hofstetter},
  \citenamefont {Neppl}, \citenamefont {Cavalieri}, \citenamefont {Komninos},
  \citenamefont {Mercouris}, \citenamefont {Nicolaides}, \citenamefont
  {Pazourek}, \citenamefont {Nagele}, \citenamefont {Feist}, \citenamefont
  {Burgd{\"o}rfer}, \citenamefont {Azzeer}, \citenamefont {Ernstorfer},
  \citenamefont {Kienberger}, \citenamefont {Kleineberg}, \citenamefont
  {Goulielmakis}, \citenamefont {Krausz},\ and\ \citenamefont
  {Yakovlev}}]{Schultze2010}%
  \BibitemOpen
  \bibfield  {author} {\bibinfo {author} {\bibfnamefont {M.}~\bibnamefont
  {Schultze}}, \bibinfo {author} {\bibfnamefont {M.}~\bibnamefont {Fie{\ss}}},
  \bibinfo {author} {\bibfnamefont {N.}~\bibnamefont {Karpowicz}}, \bibinfo
  {author} {\bibfnamefont {J.}~\bibnamefont {Gagnon}}, \bibinfo {author}
  {\bibfnamefont {M.}~\bibnamefont {Korbman}}, \bibinfo {author} {\bibfnamefont
  {M.}~\bibnamefont {Hofstetter}}, \bibinfo {author} {\bibfnamefont
  {S.}~\bibnamefont {Neppl}}, \bibinfo {author} {\bibfnamefont {A.~L.}\
  \bibnamefont {Cavalieri}}, \bibinfo {author} {\bibfnamefont {Y.}~\bibnamefont
  {Komninos}}, \bibinfo {author} {\bibfnamefont {T.}~\bibnamefont {Mercouris}},
  \bibinfo {author} {\bibfnamefont {C.~A.}\ \bibnamefont {Nicolaides}},
  \bibinfo {author} {\bibfnamefont {R.}~\bibnamefont {Pazourek}}, \bibinfo
  {author} {\bibfnamefont {S.}~\bibnamefont {Nagele}}, \bibinfo {author}
  {\bibfnamefont {J.}~\bibnamefont {Feist}}, \bibinfo {author} {\bibfnamefont
  {J.}~\bibnamefont {Burgd{\"o}rfer}}, \bibinfo {author} {\bibfnamefont
  {A.~M.}\ \bibnamefont {Azzeer}}, \bibinfo {author} {\bibfnamefont
  {R.}~\bibnamefont {Ernstorfer}}, \bibinfo {author} {\bibfnamefont
  {R.}~\bibnamefont {Kienberger}}, \bibinfo {author} {\bibfnamefont
  {U.}~\bibnamefont {Kleineberg}}, \bibinfo {author} {\bibfnamefont
  {E.}~\bibnamefont {Goulielmakis}}, \bibinfo {author} {\bibfnamefont
  {F.}~\bibnamefont {Krausz}},\ and\ \bibinfo {author} {\bibfnamefont {V.~S.}\
  \bibnamefont {Yakovlev}},\ }\bibfield  {title} {\bibinfo {title} {Delay in
  photoemission},\ }\href {https://doi.org/10.1126/science.1189401} {\bibfield
  {journal} {\bibinfo  {journal} {Science}\ }\textbf {\bibinfo {volume}
  {328}},\ \bibinfo {pages} {1658} (\bibinfo {year} {2010})}\BibitemShut
  {NoStop}%
\bibitem [{\citenamefont {Kl\"under}\ \emph {et~al.}(2011)\citenamefont
  {Kl\"under}, \citenamefont {Dahlstr\"om}, \citenamefont {Gisselbrecht},
  \citenamefont {Fordell}, \citenamefont {Swoboda}, \citenamefont {Gu\'enot},
  \citenamefont {Johnsson}, \citenamefont {Caillat}, \citenamefont
  {Mauritsson}, \citenamefont {Maquet}, \citenamefont {Ta\"{\i}eb},\ and\
  \citenamefont {L'Huillier}}]{Klunder2011}%
  \BibitemOpen
  \bibfield  {author} {\bibinfo {author} {\bibfnamefont {K.}~\bibnamefont
  {Kl\"under}}, \bibinfo {author} {\bibfnamefont {J.~M.}\ \bibnamefont
  {Dahlstr\"om}}, \bibinfo {author} {\bibfnamefont {M.}~\bibnamefont
  {Gisselbrecht}}, \bibinfo {author} {\bibfnamefont {T.}~\bibnamefont
  {Fordell}}, \bibinfo {author} {\bibfnamefont {M.}~\bibnamefont {Swoboda}},
  \bibinfo {author} {\bibfnamefont {D.}~\bibnamefont {Gu\'enot}}, \bibinfo
  {author} {\bibfnamefont {P.}~\bibnamefont {Johnsson}}, \bibinfo {author}
  {\bibfnamefont {J.}~\bibnamefont {Caillat}}, \bibinfo {author} {\bibfnamefont
  {J.}~\bibnamefont {Mauritsson}}, \bibinfo {author} {\bibfnamefont
  {A.}~\bibnamefont {Maquet}}, \bibinfo {author} {\bibfnamefont
  {R.}~\bibnamefont {Ta\"{\i}eb}},\ and\ \bibinfo {author} {\bibfnamefont
  {A.}~\bibnamefont {L'Huillier}},\ }\bibfield  {title} {\bibinfo {title}
  {Probing single-photon ionization on the attosecond time scale},\ }\href
  {https://doi.org/10.1103/PhysRevLett.106.143002} {\bibfield  {journal}
  {\bibinfo  {journal} {Phys. Rev. Lett.}\ }\textbf {\bibinfo {volume} {106}},\
  \bibinfo {pages} {143002} (\bibinfo {year} {2011})}\BibitemShut {NoStop}%
\bibitem [{\citenamefont {Froissart}\ \emph {et~al.}(1963)\citenamefont
  {Froissart}, \citenamefont {Goldberger},\ and\ \citenamefont
  {Watson}}]{Froissart1963}%
  \BibitemOpen
  \bibfield  {author} {\bibinfo {author} {\bibfnamefont {M.}~\bibnamefont
  {Froissart}}, \bibinfo {author} {\bibfnamefont {M.~L.}\ \bibnamefont
  {Goldberger}},\ and\ \bibinfo {author} {\bibfnamefont {K.~M.}\ \bibnamefont
  {Watson}},\ }\bibfield  {title} {\bibinfo {title} {Spatial separation of
  events in $s$-matrix theory},\ }\href
  {https://doi.org/10.1103/PhysRev.131.2820} {\bibfield  {journal} {\bibinfo
  {journal} {Phys. Rev.}\ }\textbf {\bibinfo {volume} {131}},\ \bibinfo {pages}
  {2820} (\bibinfo {year} {1963})}\BibitemShut {NoStop}%
\bibitem [{\citenamefont {Pazourek}\ \emph {et~al.}(2013)\citenamefont
  {Pazourek}, \citenamefont {Nagele},\ and\ \citenamefont
  {Burgdörfer}}]{Pazourek2013}%
  \BibitemOpen
  \bibfield  {author} {\bibinfo {author} {\bibfnamefont {R.}~\bibnamefont
  {Pazourek}}, \bibinfo {author} {\bibfnamefont {S.}~\bibnamefont {Nagele}},\
  and\ \bibinfo {author} {\bibfnamefont {J.}~\bibnamefont {Burgdörfer}},\
  }\bibfield  {title} {\bibinfo {title} {Time-resolved photoemission on the
  attosecond scale: opportunities and challenges},\ }\href
  {https://doi.org/10.1039/C3FD00004D} {\bibfield  {journal} {\bibinfo
  {journal} {Faraday Discuss.}\ }\textbf {\bibinfo {volume} {163}},\ \bibinfo
  {pages} {353} (\bibinfo {year} {2013})}\BibitemShut {NoStop}%
\bibitem [{\citenamefont {Nagele}\ \emph {et~al.}(2011)\citenamefont {Nagele},
  \citenamefont {Pazourek}, \citenamefont {Feist}, \citenamefont
  {Doblhoff-Dier}, \citenamefont {Lemell}, \citenamefont {T{\H{o}}k{\'{e}}si},\
  and\ \citenamefont {Burgdörfer}}]{Nagele2011}%
  \BibitemOpen
  \bibfield  {author} {\bibinfo {author} {\bibfnamefont {S.}~\bibnamefont
  {Nagele}}, \bibinfo {author} {\bibfnamefont {R.}~\bibnamefont {Pazourek}},
  \bibinfo {author} {\bibfnamefont {J.}~\bibnamefont {Feist}}, \bibinfo
  {author} {\bibfnamefont {K.}~\bibnamefont {Doblhoff-Dier}}, \bibinfo {author}
  {\bibfnamefont {C.}~\bibnamefont {Lemell}}, \bibinfo {author} {\bibfnamefont
  {K.}~\bibnamefont {T{\H{o}}k{\'{e}}si}},\ and\ \bibinfo {author}
  {\bibfnamefont {J.}~\bibnamefont {Burgdörfer}},\ }\bibfield  {title}
  {\bibinfo {title} {Time-resolved photoemission by attosecond streaking:
  extraction of time information},\ }\href
  {https://doi.org/10.1088/0953-4075/44/8/081001} {\bibfield  {journal}
  {\bibinfo  {journal} {Journal of Physics B: Atomic, Molecular and Optical
  Physics}\ }\textbf {\bibinfo {volume} {44}},\ \bibinfo {pages} {081001}
  (\bibinfo {year} {2011})}\BibitemShut {NoStop}%
\bibitem [{\citenamefont {Dahlström}\ \emph {et~al.}(2013)\citenamefont
  {Dahlström}, \citenamefont {Guénot}, \citenamefont {Klünder},
  \citenamefont {Gisselbrecht}, \citenamefont {Mauritsson}, \citenamefont
  {L’Huillier}, \citenamefont {Maquet},\ and\ \citenamefont
  {Taïeb}}]{Dahlstrom2013}%
  \BibitemOpen
  \bibfield  {author} {\bibinfo {author} {\bibfnamefont {J.}~\bibnamefont
  {Dahlström}}, \bibinfo {author} {\bibfnamefont {D.}~\bibnamefont {Guénot}},
  \bibinfo {author} {\bibfnamefont {K.}~\bibnamefont {Klünder}}, \bibinfo
  {author} {\bibfnamefont {M.}~\bibnamefont {Gisselbrecht}}, \bibinfo {author}
  {\bibfnamefont {J.}~\bibnamefont {Mauritsson}}, \bibinfo {author}
  {\bibfnamefont {A.}~\bibnamefont {L’Huillier}}, \bibinfo {author}
  {\bibfnamefont {A.}~\bibnamefont {Maquet}},\ and\ \bibinfo {author}
  {\bibfnamefont {R.}~\bibnamefont {Taïeb}},\ }\bibfield  {title} {\bibinfo
  {title} {Theory of attosecond delays in laser-assisted photoionization},\
  }\href {https://doi.org/https://doi.org/10.1016/j.chemphys.2012.01.017}
  {\bibfield  {journal} {\bibinfo  {journal} {Chemical Physics}\ }\textbf
  {\bibinfo {volume} {414}},\ \bibinfo {pages} {53} (\bibinfo {year} {2013})},\
  \bibinfo {note} {attosecond spectroscopy}\BibitemShut {NoStop}%
\bibitem [{\citenamefont {Palatchi}\ \emph {et~al.}(2014)\citenamefont
  {Palatchi}, \citenamefont {Dahlström}, \citenamefont {Kheifets},
  \citenamefont {Ivanov}, \citenamefont {Canaday}, \citenamefont {Agostini},\
  and\ \citenamefont {DiMauro}}]{Palatchi2014}%
  \BibitemOpen
  \bibfield  {author} {\bibinfo {author} {\bibfnamefont {C.}~\bibnamefont
  {Palatchi}}, \bibinfo {author} {\bibfnamefont {J.~M.}\ \bibnamefont
  {Dahlström}}, \bibinfo {author} {\bibfnamefont {A.~S.}\ \bibnamefont
  {Kheifets}}, \bibinfo {author} {\bibfnamefont {I.~A.}\ \bibnamefont
  {Ivanov}}, \bibinfo {author} {\bibfnamefont {D.~M.}\ \bibnamefont {Canaday}},
  \bibinfo {author} {\bibfnamefont {P.}~\bibnamefont {Agostini}},\ and\
  \bibinfo {author} {\bibfnamefont {L.~F.}\ \bibnamefont {DiMauro}},\
  }\bibfield  {title} {\bibinfo {title} {Atomic delay in helium, neon, argon
  and krypton},\ }\href {https://doi.org/10.1088/0953-4075/47/24/245003}
  {\bibfield  {journal} {\bibinfo  {journal} {Journal of Physics B: Atomic,
  Molecular and Optical Physics}\ }\textbf {\bibinfo {volume} {47}},\ \bibinfo
  {pages} {245003} (\bibinfo {year} {2014})}\BibitemShut {NoStop}%
\bibitem [{\citenamefont {Gu{\'{e}}not}\ \emph {et~al.}(2014)\citenamefont
  {Gu{\'{e}}not}, \citenamefont {Kroon}, \citenamefont {Balogh}, \citenamefont
  {Larsen}, \citenamefont {Kotur}, \citenamefont {Miranda}, \citenamefont
  {Fordell}, \citenamefont {Johnsson}, \citenamefont {Mauritsson},
  \citenamefont {Gisselbrecht}, \citenamefont {Varj{\`{u}}}, \citenamefont
  {Arnold}, \citenamefont {Carette}, \citenamefont {Kheifets}, \citenamefont
  {Lindroth}, \citenamefont {L'Huillier},\ and\ \citenamefont
  {Dahlström}}]{Guenot2014}%
  \BibitemOpen
  \bibfield  {author} {\bibinfo {author} {\bibfnamefont {D.}~\bibnamefont
  {Gu{\'{e}}not}}, \bibinfo {author} {\bibfnamefont {D.}~\bibnamefont {Kroon}},
  \bibinfo {author} {\bibfnamefont {E.}~\bibnamefont {Balogh}}, \bibinfo
  {author} {\bibfnamefont {E.~W.}\ \bibnamefont {Larsen}}, \bibinfo {author}
  {\bibfnamefont {M.}~\bibnamefont {Kotur}}, \bibinfo {author} {\bibfnamefont
  {M.}~\bibnamefont {Miranda}}, \bibinfo {author} {\bibfnamefont
  {T.}~\bibnamefont {Fordell}}, \bibinfo {author} {\bibfnamefont
  {P.}~\bibnamefont {Johnsson}}, \bibinfo {author} {\bibfnamefont
  {J.}~\bibnamefont {Mauritsson}}, \bibinfo {author} {\bibfnamefont
  {M.}~\bibnamefont {Gisselbrecht}}, \bibinfo {author} {\bibfnamefont
  {K.}~\bibnamefont {Varj{\`{u}}}}, \bibinfo {author} {\bibfnamefont {C.~L.}\
  \bibnamefont {Arnold}}, \bibinfo {author} {\bibfnamefont {T.}~\bibnamefont
  {Carette}}, \bibinfo {author} {\bibfnamefont {A.~S.}\ \bibnamefont
  {Kheifets}}, \bibinfo {author} {\bibfnamefont {E.}~\bibnamefont {Lindroth}},
  \bibinfo {author} {\bibfnamefont {A.}~\bibnamefont {L'Huillier}},\ and\
  \bibinfo {author} {\bibfnamefont {J.~M.}\ \bibnamefont {Dahlström}},\
  }\bibfield  {title} {\bibinfo {title} {Measurements of relative photoemission
  time delays in noble gas atoms},\ }\href
  {https://doi.org/10.1088/0953-4075/47/24/245602} {\bibfield  {journal}
  {\bibinfo  {journal} {Journal of Physics B: Atomic, Molecular and Optical
  Physics}\ }\textbf {\bibinfo {volume} {47}},\ \bibinfo {pages} {245602}
  (\bibinfo {year} {2014})}\BibitemShut {NoStop}%
\bibitem [{\citenamefont {Huppert}\ \emph {et~al.}(2016)\citenamefont
  {Huppert}, \citenamefont {Jordan}, \citenamefont {Baykusheva}, \citenamefont
  {von Conta},\ and\ \citenamefont {W\"orner}}]{Huppert2016}%
  \BibitemOpen
  \bibfield  {author} {\bibinfo {author} {\bibfnamefont {M.}~\bibnamefont
  {Huppert}}, \bibinfo {author} {\bibfnamefont {I.}~\bibnamefont {Jordan}},
  \bibinfo {author} {\bibfnamefont {D.}~\bibnamefont {Baykusheva}}, \bibinfo
  {author} {\bibfnamefont {A.}~\bibnamefont {von Conta}},\ and\ \bibinfo
  {author} {\bibfnamefont {H.~J.}\ \bibnamefont {W\"orner}},\ }\bibfield
  {title} {\bibinfo {title} {Attosecond delays in molecular photoionization},\
  }\href {https://doi.org/10.1103/PhysRevLett.117.093001} {\bibfield  {journal}
  {\bibinfo  {journal} {Phys. Rev. Lett.}\ }\textbf {\bibinfo {volume} {117}},\
  \bibinfo {pages} {093001} (\bibinfo {year} {2016})}\BibitemShut {NoStop}%
\bibitem [{\citenamefont {Baykusheva}\ and\ \citenamefont
  {Wörner}(2017)}]{Baykusheva2017}%
  \BibitemOpen
  \bibfield  {author} {\bibinfo {author} {\bibfnamefont {D.}~\bibnamefont
  {Baykusheva}}\ and\ \bibinfo {author} {\bibfnamefont {H.~J.}\ \bibnamefont
  {Wörner}},\ }\bibfield  {title} {\bibinfo {title} {Theory of attosecond
  delays in molecular photoionization},\ }\href
  {https://doi.org/10.1063/1.4977933} {\bibfield  {journal} {\bibinfo
  {journal} {The Journal of Chemical Physics}\ }\textbf {\bibinfo {volume}
  {146}},\ \bibinfo {pages} {124306} (\bibinfo {year} {2017})},\ \Eprint
  {https://arxiv.org/abs/https://doi.org/10.1063/1.4977933}
  {https://doi.org/10.1063/1.4977933} \BibitemShut {NoStop}%
\bibitem [{\citenamefont {Heuser}\ \emph {et~al.}(2016)\citenamefont {Heuser},
  \citenamefont {Jim\'enez~Gal\'an}, \citenamefont {Cirelli}, \citenamefont
  {Marante}, \citenamefont {Sabbar}, \citenamefont {Boge}, \citenamefont
  {Lucchini}, \citenamefont {Gallmann}, \citenamefont {Ivanov}, \citenamefont
  {Kheifets}, \citenamefont {Dahlstr\"om}, \citenamefont {Lindroth},
  \citenamefont {Argenti}, \citenamefont {Mart\'{\i}n},\ and\ \citenamefont
  {Keller}}]{Heuser2016}%
  \BibitemOpen
  \bibfield  {author} {\bibinfo {author} {\bibfnamefont {S.}~\bibnamefont
  {Heuser}}, \bibinfo {author} {\bibfnamefont {A.}~\bibnamefont
  {Jim\'enez~Gal\'an}}, \bibinfo {author} {\bibfnamefont {C.}~\bibnamefont
  {Cirelli}}, \bibinfo {author} {\bibfnamefont {C.}~\bibnamefont {Marante}},
  \bibinfo {author} {\bibfnamefont {M.}~\bibnamefont {Sabbar}}, \bibinfo
  {author} {\bibfnamefont {R.}~\bibnamefont {Boge}}, \bibinfo {author}
  {\bibfnamefont {M.}~\bibnamefont {Lucchini}}, \bibinfo {author}
  {\bibfnamefont {L.}~\bibnamefont {Gallmann}}, \bibinfo {author}
  {\bibfnamefont {I.}~\bibnamefont {Ivanov}}, \bibinfo {author} {\bibfnamefont
  {A.~S.}\ \bibnamefont {Kheifets}}, \bibinfo {author} {\bibfnamefont {J.~M.}\
  \bibnamefont {Dahlstr\"om}}, \bibinfo {author} {\bibfnamefont
  {E.}~\bibnamefont {Lindroth}}, \bibinfo {author} {\bibfnamefont
  {L.}~\bibnamefont {Argenti}}, \bibinfo {author} {\bibfnamefont
  {F.}~\bibnamefont {Mart\'{\i}n}},\ and\ \bibinfo {author} {\bibfnamefont
  {U.}~\bibnamefont {Keller}},\ }\bibfield  {title} {\bibinfo {title} {Angular
  dependence of photoemission time delay in helium},\ }\href
  {https://doi.org/10.1103/PhysRevA.94.063409} {\bibfield  {journal} {\bibinfo
  {journal} {Phys. Rev. A}\ }\textbf {\bibinfo {volume} {94}},\ \bibinfo
  {pages} {063409} (\bibinfo {year} {2016})}\BibitemShut {NoStop}%
\bibitem [{\citenamefont {Cirelli}\ \emph {et~al.}(2018)\citenamefont
  {Cirelli}, \citenamefont {Marante}, \citenamefont {Heuser}, \citenamefont
  {Petersson}, \citenamefont {Gal{\'a}n}, \citenamefont {Argenti},
  \citenamefont {Zhong}, \citenamefont {Busto}, \citenamefont {Isinger},
  \citenamefont {Nandi}, \citenamefont {Maclot}, \citenamefont {Rading},
  \citenamefont {Johnsson}, \citenamefont {Gisselbrecht}, \citenamefont
  {Lucchini}, \citenamefont {Gallmann}, \citenamefont {Dahlstr{\"o}m},
  \citenamefont {Lindroth}, \citenamefont {L'Huillier}, \citenamefont
  {Mart{\'i}n},\ and\ \citenamefont {Keller}}]{Cirelli2018}%
  \BibitemOpen
  \bibfield  {author} {\bibinfo {author} {\bibfnamefont {C.}~\bibnamefont
  {Cirelli}}, \bibinfo {author} {\bibfnamefont {C.}~\bibnamefont {Marante}},
  \bibinfo {author} {\bibfnamefont {S.}~\bibnamefont {Heuser}}, \bibinfo
  {author} {\bibfnamefont {C.~L.~M.}\ \bibnamefont {Petersson}}, \bibinfo
  {author} {\bibfnamefont {{\'A}.~J.}\ \bibnamefont {Gal{\'a}n}}, \bibinfo
  {author} {\bibfnamefont {L.}~\bibnamefont {Argenti}}, \bibinfo {author}
  {\bibfnamefont {S.}~\bibnamefont {Zhong}}, \bibinfo {author} {\bibfnamefont
  {D.}~\bibnamefont {Busto}}, \bibinfo {author} {\bibfnamefont
  {M.}~\bibnamefont {Isinger}}, \bibinfo {author} {\bibfnamefont
  {S.}~\bibnamefont {Nandi}}, \bibinfo {author} {\bibfnamefont
  {S.}~\bibnamefont {Maclot}}, \bibinfo {author} {\bibfnamefont
  {L.}~\bibnamefont {Rading}}, \bibinfo {author} {\bibfnamefont
  {P.}~\bibnamefont {Johnsson}}, \bibinfo {author} {\bibfnamefont
  {M.}~\bibnamefont {Gisselbrecht}}, \bibinfo {author} {\bibfnamefont
  {M.}~\bibnamefont {Lucchini}}, \bibinfo {author} {\bibfnamefont
  {L.}~\bibnamefont {Gallmann}}, \bibinfo {author} {\bibfnamefont {J.~M.}\
  \bibnamefont {Dahlstr{\"o}m}}, \bibinfo {author} {\bibfnamefont
  {E.}~\bibnamefont {Lindroth}}, \bibinfo {author} {\bibfnamefont
  {A.}~\bibnamefont {L'Huillier}}, \bibinfo {author} {\bibfnamefont
  {F.}~\bibnamefont {Mart{\'i}n}},\ and\ \bibinfo {author} {\bibfnamefont
  {U.}~\bibnamefont {Keller}},\ }\bibfield  {title} {\bibinfo {title}
  {Anisotropic photoemission time delays close to a fano resonance},\ }\href
  {https://doi.org/10.1038/s41467-018-03009-1} {\bibfield  {journal} {\bibinfo
  {journal} {Nature Communications}\ }\textbf {\bibinfo {volume} {9}},\
  \bibinfo {pages} {955} (\bibinfo {year} {2018})}\BibitemShut {NoStop}%
\bibitem [{\citenamefont {Busto}\ \emph {et~al.}(2019)\citenamefont {Busto},
  \citenamefont {Vinbladh}, \citenamefont {Zhong}, \citenamefont {Isinger},
  \citenamefont {Nandi}, \citenamefont {Maclot}, \citenamefont {Johnsson},
  \citenamefont {Gisselbrecht}, \citenamefont {L'Huillier}, \citenamefont
  {Lindroth},\ and\ \citenamefont {Dahlstr\"om}}]{Busto2019}%
  \BibitemOpen
  \bibfield  {author} {\bibinfo {author} {\bibfnamefont {D.}~\bibnamefont
  {Busto}}, \bibinfo {author} {\bibfnamefont {J.}~\bibnamefont {Vinbladh}},
  \bibinfo {author} {\bibfnamefont {S.}~\bibnamefont {Zhong}}, \bibinfo
  {author} {\bibfnamefont {M.}~\bibnamefont {Isinger}}, \bibinfo {author}
  {\bibfnamefont {S.}~\bibnamefont {Nandi}}, \bibinfo {author} {\bibfnamefont
  {S.}~\bibnamefont {Maclot}}, \bibinfo {author} {\bibfnamefont
  {P.}~\bibnamefont {Johnsson}}, \bibinfo {author} {\bibfnamefont
  {M.}~\bibnamefont {Gisselbrecht}}, \bibinfo {author} {\bibfnamefont
  {A.}~\bibnamefont {L'Huillier}}, \bibinfo {author} {\bibfnamefont
  {E.}~\bibnamefont {Lindroth}},\ and\ \bibinfo {author} {\bibfnamefont
  {J.~M.}\ \bibnamefont {Dahlstr\"om}},\ }\bibfield  {title} {\bibinfo {title}
  {Fano's propensity rule in angle-resolved attosecond pump-probe
  photoionization},\ }\href {https://doi.org/10.1103/PhysRevLett.123.133201}
  {\bibfield  {journal} {\bibinfo  {journal} {Phys. Rev. Lett.}\ }\textbf
  {\bibinfo {volume} {123}},\ \bibinfo {pages} {133201} (\bibinfo {year}
  {2019})}\BibitemShut {NoStop}%
\bibitem [{\citenamefont {Autuori}\ \emph {et~al.}(2022)\citenamefont
  {Autuori}, \citenamefont {Platzer}, \citenamefont {Lejman}, \citenamefont
  {Gallician}, \citenamefont {Maëder}, \citenamefont {Covolo}, \citenamefont
  {Bosse}, \citenamefont {Dalui}, \citenamefont {Bresteau}, \citenamefont
  {Hergott}, \citenamefont {Tcherbakoff}, \citenamefont {Marroux},
  \citenamefont {Loriot}, \citenamefont {Lépine}, \citenamefont {Poisson},
  \citenamefont {Taïeb}, \citenamefont {Caillat},\ and\ \citenamefont
  {Salières}}]{Autuori2022}%
  \BibitemOpen
  \bibfield  {author} {\bibinfo {author} {\bibfnamefont {A.}~\bibnamefont
  {Autuori}}, \bibinfo {author} {\bibfnamefont {D.}~\bibnamefont {Platzer}},
  \bibinfo {author} {\bibfnamefont {M.}~\bibnamefont {Lejman}}, \bibinfo
  {author} {\bibfnamefont {G.}~\bibnamefont {Gallician}}, \bibinfo {author}
  {\bibfnamefont {L.}~\bibnamefont {Maëder}}, \bibinfo {author} {\bibfnamefont
  {A.}~\bibnamefont {Covolo}}, \bibinfo {author} {\bibfnamefont
  {L.}~\bibnamefont {Bosse}}, \bibinfo {author} {\bibfnamefont
  {M.}~\bibnamefont {Dalui}}, \bibinfo {author} {\bibfnamefont
  {D.}~\bibnamefont {Bresteau}}, \bibinfo {author} {\bibfnamefont {J.-F.}\
  \bibnamefont {Hergott}}, \bibinfo {author} {\bibfnamefont {O.}~\bibnamefont
  {Tcherbakoff}}, \bibinfo {author} {\bibfnamefont {H.~J.~B.}\ \bibnamefont
  {Marroux}}, \bibinfo {author} {\bibfnamefont {V.}~\bibnamefont {Loriot}},
  \bibinfo {author} {\bibfnamefont {F.}~\bibnamefont {Lépine}}, \bibinfo
  {author} {\bibfnamefont {L.}~\bibnamefont {Poisson}}, \bibinfo {author}
  {\bibfnamefont {R.}~\bibnamefont {Taïeb}}, \bibinfo {author} {\bibfnamefont
  {J.}~\bibnamefont {Caillat}},\ and\ \bibinfo {author} {\bibfnamefont
  {P.}~\bibnamefont {Salières}},\ }\bibfield  {title} {\bibinfo {title}
  {Anisotropic dynamics of two-photon ionization: An attosecond movie of
  photoemission},\ }\href {https://doi.org/10.1126/sciadv.abl7594} {\bibfield
  {journal} {\bibinfo  {journal} {Science Advances}\ }\textbf {\bibinfo
  {volume} {8}},\ \bibinfo {pages} {eabl7594} (\bibinfo {year} {2022})},\
  \Eprint
  {https://arxiv.org/abs/https://www.science.org/doi/pdf/10.1126/sciadv.abl7594}
  {https://www.science.org/doi/pdf/10.1126/sciadv.abl7594} \BibitemShut
  {NoStop}%
\bibitem [{\citenamefont {Vos}\ \emph {et~al.}(2018)\citenamefont {Vos},
  \citenamefont {Cattaneo}, \citenamefont {Patchkovskii}, \citenamefont
  {Zimmermann}, \citenamefont {Cirelli}, \citenamefont {Lucchini},
  \citenamefont {Kheifets}, \citenamefont {Landsman},\ and\ \citenamefont
  {Keller}}]{Vos2018}%
  \BibitemOpen
  \bibfield  {author} {\bibinfo {author} {\bibfnamefont {J.}~\bibnamefont
  {Vos}}, \bibinfo {author} {\bibfnamefont {L.}~\bibnamefont {Cattaneo}},
  \bibinfo {author} {\bibfnamefont {S.}~\bibnamefont {Patchkovskii}}, \bibinfo
  {author} {\bibfnamefont {T.}~\bibnamefont {Zimmermann}}, \bibinfo {author}
  {\bibfnamefont {C.}~\bibnamefont {Cirelli}}, \bibinfo {author} {\bibfnamefont
  {M.}~\bibnamefont {Lucchini}}, \bibinfo {author} {\bibfnamefont
  {A.}~\bibnamefont {Kheifets}}, \bibinfo {author} {\bibfnamefont {A.~S.}\
  \bibnamefont {Landsman}},\ and\ \bibinfo {author} {\bibfnamefont
  {U.}~\bibnamefont {Keller}},\ }\bibfield  {title} {\bibinfo {title}
  {Orientation-dependent stereo wigner time delay and electron localization in
  a small molecule},\ }\href {https://doi.org/10.1126/science.aao4731}
  {\bibfield  {journal} {\bibinfo  {journal} {Science}\ }\textbf {\bibinfo
  {volume} {360}},\ \bibinfo {pages} {1326} (\bibinfo {year} {2018})},\ \Eprint
  {https://arxiv.org/abs/https://www.science.org/doi/pdf/10.1126/science.aao4731}
  {https://www.science.org/doi/pdf/10.1126/science.aao4731} \BibitemShut
  {NoStop}%
\bibitem [{\citenamefont {Nandi}\ \emph {et~al.}(2020)\citenamefont {Nandi},
  \citenamefont {Plésiat}, \citenamefont {Zhong}, \citenamefont {Palacios},
  \citenamefont {Busto}, \citenamefont {Isinger}, \citenamefont {Neoričić},
  \citenamefont {Arnold}, \citenamefont {Squibb}, \citenamefont {Feifel},
  \citenamefont {Decleva}, \citenamefont {L’Huillier}, \citenamefont
  {Martín},\ and\ \citenamefont {Gisselbrecht}}]{Nandi2020}%
  \BibitemOpen
  \bibfield  {author} {\bibinfo {author} {\bibfnamefont {S.}~\bibnamefont
  {Nandi}}, \bibinfo {author} {\bibfnamefont {E.}~\bibnamefont {Plésiat}},
  \bibinfo {author} {\bibfnamefont {S.}~\bibnamefont {Zhong}}, \bibinfo
  {author} {\bibfnamefont {A.}~\bibnamefont {Palacios}}, \bibinfo {author}
  {\bibfnamefont {D.}~\bibnamefont {Busto}}, \bibinfo {author} {\bibfnamefont
  {M.}~\bibnamefont {Isinger}}, \bibinfo {author} {\bibfnamefont
  {L.}~\bibnamefont {Neoričić}}, \bibinfo {author} {\bibfnamefont {C.~L.}\
  \bibnamefont {Arnold}}, \bibinfo {author} {\bibfnamefont {R.~J.}\
  \bibnamefont {Squibb}}, \bibinfo {author} {\bibfnamefont {R.}~\bibnamefont
  {Feifel}}, \bibinfo {author} {\bibfnamefont {P.}~\bibnamefont {Decleva}},
  \bibinfo {author} {\bibfnamefont {A.}~\bibnamefont {L’Huillier}}, \bibinfo
  {author} {\bibfnamefont {F.}~\bibnamefont {Martín}},\ and\ \bibinfo {author}
  {\bibfnamefont {M.}~\bibnamefont {Gisselbrecht}},\ }\bibfield  {title}
  {\bibinfo {title} {Attosecond timing of electron emission from a molecular
  shape resonance},\ }\href {https://doi.org/10.1126/sciadv.aba7762} {\bibfield
   {journal} {\bibinfo  {journal} {Science Advances}\ }\textbf {\bibinfo
  {volume} {6}},\ \bibinfo {pages} {eaba7762} (\bibinfo {year} {2020})},\
  \Eprint
  {https://arxiv.org/abs/https://www.science.org/doi/pdf/10.1126/sciadv.aba7762}
  {https://www.science.org/doi/pdf/10.1126/sciadv.aba7762} \BibitemShut
  {NoStop}%
\bibitem [{\citenamefont {Holzmeier}\ \emph {et~al.}(2021)\citenamefont
  {Holzmeier}, \citenamefont {Joseph}, \citenamefont {Houver}, \citenamefont
  {Lebech}, \citenamefont {Dowek},\ and\ \citenamefont
  {Lucchese}}]{Holzmeier2021}%
  \BibitemOpen
  \bibfield  {author} {\bibinfo {author} {\bibfnamefont {F.}~\bibnamefont
  {Holzmeier}}, \bibinfo {author} {\bibfnamefont {J.}~\bibnamefont {Joseph}},
  \bibinfo {author} {\bibfnamefont {J.~C.}\ \bibnamefont {Houver}}, \bibinfo
  {author} {\bibfnamefont {M.}~\bibnamefont {Lebech}}, \bibinfo {author}
  {\bibfnamefont {D.}~\bibnamefont {Dowek}},\ and\ \bibinfo {author}
  {\bibfnamefont {R.~R.}\ \bibnamefont {Lucchese}},\ }\bibfield  {title}
  {\bibinfo {title} {Influence of shape resonances on the angular dependence of
  molecular photoionization delays},\ }\href
  {https://doi.org/10.1038/s41467-021-27360-y} {\bibfield  {journal} {\bibinfo
  {journal} {Nature Communications}\ }\textbf {\bibinfo {volume} {12}},\
  \bibinfo {pages} {7343} (\bibinfo {year} {2021})}\BibitemShut {NoStop}%
\bibitem [{\citenamefont {Ahmadi}\ \emph {et~al.}(2022)\citenamefont {Ahmadi},
  \citenamefont {Pl{\'e}siat}, \citenamefont {Moioli}, \citenamefont
  {Frassetto}, \citenamefont {Poletto}, \citenamefont {Decleva}, \citenamefont
  {Schr{\"o}ter}, \citenamefont {Pfeifer}, \citenamefont {Moshammer},
  \citenamefont {Palacios}, \citenamefont {Martin},\ and\ \citenamefont
  {Sansone}}]{Ahmadi2022}%
  \BibitemOpen
  \bibfield  {author} {\bibinfo {author} {\bibfnamefont {H.}~\bibnamefont
  {Ahmadi}}, \bibinfo {author} {\bibfnamefont {E.}~\bibnamefont {Pl{\'e}siat}},
  \bibinfo {author} {\bibfnamefont {M.}~\bibnamefont {Moioli}}, \bibinfo
  {author} {\bibfnamefont {F.}~\bibnamefont {Frassetto}}, \bibinfo {author}
  {\bibfnamefont {L.}~\bibnamefont {Poletto}}, \bibinfo {author} {\bibfnamefont
  {P.}~\bibnamefont {Decleva}}, \bibinfo {author} {\bibfnamefont {C.~D.}\
  \bibnamefont {Schr{\"o}ter}}, \bibinfo {author} {\bibfnamefont
  {T.}~\bibnamefont {Pfeifer}}, \bibinfo {author} {\bibfnamefont
  {R.}~\bibnamefont {Moshammer}}, \bibinfo {author} {\bibfnamefont
  {A.}~\bibnamefont {Palacios}}, \bibinfo {author} {\bibfnamefont
  {F.}~\bibnamefont {Martin}},\ and\ \bibinfo {author} {\bibfnamefont
  {G.}~\bibnamefont {Sansone}},\ }\bibfield  {title} {\bibinfo {title}
  {Attosecond photoionisation time delays reveal the anisotropy of the
  molecular potential in the recoil frame},\ }\href
  {https://doi.org/10.1038/s41467-022-28783-x} {\bibfield  {journal} {\bibinfo
  {journal} {Nature Communications}\ }\textbf {\bibinfo {volume} {13}},\
  \bibinfo {pages} {1242} (\bibinfo {year} {2022})}\BibitemShut {NoStop}%
\bibitem [{\citenamefont {O'Keeffe}\ \emph {et~al.}(2004)\citenamefont
  {O'Keeffe}, \citenamefont {L\'opez-Martens}, \citenamefont {Mauritsson},
  \citenamefont {Johansson}, \citenamefont {L'Huillier}, \citenamefont
  {V\'eniard}, \citenamefont {Ta\"{\i}eb}, \citenamefont {Maquet},\ and\
  \citenamefont {Meyer}}]{OKeeffe2004}%
  \BibitemOpen
  \bibfield  {author} {\bibinfo {author} {\bibfnamefont {P.}~\bibnamefont
  {O'Keeffe}}, \bibinfo {author} {\bibfnamefont {R.}~\bibnamefont
  {L\'opez-Martens}}, \bibinfo {author} {\bibfnamefont {J.}~\bibnamefont
  {Mauritsson}}, \bibinfo {author} {\bibfnamefont {A.}~\bibnamefont
  {Johansson}}, \bibinfo {author} {\bibfnamefont {A.}~\bibnamefont
  {L'Huillier}}, \bibinfo {author} {\bibfnamefont {V.}~\bibnamefont
  {V\'eniard}}, \bibinfo {author} {\bibfnamefont {R.}~\bibnamefont
  {Ta\"{\i}eb}}, \bibinfo {author} {\bibfnamefont {A.}~\bibnamefont {Maquet}},\
  and\ \bibinfo {author} {\bibfnamefont {M.}~\bibnamefont {Meyer}},\ }\bibfield
   {title} {\bibinfo {title} {Polarization effects in two-photon nonresonant
  ionization of argon with extreme-ultraviolet and infrared femtosecond
  pulses},\ }\href {https://doi.org/10.1103/PhysRevA.69.051401} {\bibfield
  {journal} {\bibinfo  {journal} {Phys. Rev. A}\ }\textbf {\bibinfo {volume}
  {69}},\ \bibinfo {pages} {051401} (\bibinfo {year} {2004})}\BibitemShut
  {NoStop}%
\bibitem [{\citenamefont {Meyer}\ \emph {et~al.}(2008)\citenamefont {Meyer},
  \citenamefont {Cubaynes}, \citenamefont {Glijer}, \citenamefont {Dardis},
  \citenamefont {Hayden}, \citenamefont {Hough}, \citenamefont {Richardson},
  \citenamefont {Kennedy}, \citenamefont {Costello}, \citenamefont {Radcliffe},
  \citenamefont {D\"usterer}, \citenamefont {Azima}, \citenamefont {Li},
  \citenamefont {Redlin}, \citenamefont {Feldhaus}, \citenamefont {Ta\"{\i}eb},
  \citenamefont {Maquet}, \citenamefont {Grum-Grzhimailo}, \citenamefont
  {Gryzlova},\ and\ \citenamefont {Strakhova}}]{Meyer2008}%
  \BibitemOpen
  \bibfield  {author} {\bibinfo {author} {\bibfnamefont {M.}~\bibnamefont
  {Meyer}}, \bibinfo {author} {\bibfnamefont {D.}~\bibnamefont {Cubaynes}},
  \bibinfo {author} {\bibfnamefont {D.}~\bibnamefont {Glijer}}, \bibinfo
  {author} {\bibfnamefont {J.}~\bibnamefont {Dardis}}, \bibinfo {author}
  {\bibfnamefont {P.}~\bibnamefont {Hayden}}, \bibinfo {author} {\bibfnamefont
  {P.}~\bibnamefont {Hough}}, \bibinfo {author} {\bibfnamefont
  {V.}~\bibnamefont {Richardson}}, \bibinfo {author} {\bibfnamefont {E.~T.}\
  \bibnamefont {Kennedy}}, \bibinfo {author} {\bibfnamefont {J.~T.}\
  \bibnamefont {Costello}}, \bibinfo {author} {\bibfnamefont {P.}~\bibnamefont
  {Radcliffe}}, \bibinfo {author} {\bibfnamefont {S.}~\bibnamefont
  {D\"usterer}}, \bibinfo {author} {\bibfnamefont {A.}~\bibnamefont {Azima}},
  \bibinfo {author} {\bibfnamefont {W.~B.}\ \bibnamefont {Li}}, \bibinfo
  {author} {\bibfnamefont {H.}~\bibnamefont {Redlin}}, \bibinfo {author}
  {\bibfnamefont {J.}~\bibnamefont {Feldhaus}}, \bibinfo {author}
  {\bibfnamefont {R.}~\bibnamefont {Ta\"{\i}eb}}, \bibinfo {author}
  {\bibfnamefont {A.}~\bibnamefont {Maquet}}, \bibinfo {author} {\bibfnamefont
  {A.~N.}\ \bibnamefont {Grum-Grzhimailo}}, \bibinfo {author} {\bibfnamefont
  {E.~V.}\ \bibnamefont {Gryzlova}},\ and\ \bibinfo {author} {\bibfnamefont
  {S.~I.}\ \bibnamefont {Strakhova}},\ }\bibfield  {title} {\bibinfo {title}
  {Polarization control in two-color above-threshold ionization of atomic
  helium},\ }\href {https://doi.org/10.1103/PhysRevLett.101.193002} {\bibfield
  {journal} {\bibinfo  {journal} {Phys. Rev. Lett.}\ }\textbf {\bibinfo
  {volume} {101}},\ \bibinfo {pages} {193002} (\bibinfo {year}
  {2008})}\BibitemShut {NoStop}%
\bibitem [{\citenamefont {Jiang}\ \emph {et~al.}(2022)\citenamefont {Jiang},
  \citenamefont {Armstrong}, \citenamefont {Tong}, \citenamefont {Xu},
  \citenamefont {Zuo}, \citenamefont {Qiang}, \citenamefont {Lu}, \citenamefont
  {Clarke}, \citenamefont {Benda}, \citenamefont {Fleischer}, \citenamefont
  {Ni}, \citenamefont {Ueda}, \citenamefont {van~der Hart}, \citenamefont
  {Brown}, \citenamefont {Gong},\ and\ \citenamefont {Wu}}]{Jiang2022}%
  \BibitemOpen
  \bibfield  {author} {\bibinfo {author} {\bibfnamefont {W.}~\bibnamefont
  {Jiang}}, \bibinfo {author} {\bibfnamefont {G.~S.~J.}\ \bibnamefont
  {Armstrong}}, \bibinfo {author} {\bibfnamefont {J.}~\bibnamefont {Tong}},
  \bibinfo {author} {\bibfnamefont {Y.}~\bibnamefont {Xu}}, \bibinfo {author}
  {\bibfnamefont {Z.}~\bibnamefont {Zuo}}, \bibinfo {author} {\bibfnamefont
  {J.}~\bibnamefont {Qiang}}, \bibinfo {author} {\bibfnamefont
  {P.}~\bibnamefont {Lu}}, \bibinfo {author} {\bibfnamefont {D.~D.~A.}\
  \bibnamefont {Clarke}}, \bibinfo {author} {\bibfnamefont {J.}~\bibnamefont
  {Benda}}, \bibinfo {author} {\bibfnamefont {A.}~\bibnamefont {Fleischer}},
  \bibinfo {author} {\bibfnamefont {H.}~\bibnamefont {Ni}}, \bibinfo {author}
  {\bibfnamefont {K.}~\bibnamefont {Ueda}}, \bibinfo {author} {\bibfnamefont
  {H.~W.}\ \bibnamefont {van~der Hart}}, \bibinfo {author} {\bibfnamefont
  {A.~C.}\ \bibnamefont {Brown}}, \bibinfo {author} {\bibfnamefont
  {X.}~\bibnamefont {Gong}},\ and\ \bibinfo {author} {\bibfnamefont
  {J.}~\bibnamefont {Wu}},\ }\bibfield  {title} {\bibinfo {title} {Atomic
  partial wave meter by attosecond coincidence metrology},\ }\href
  {https://doi.org/10.1038/s41467-022-32753-8} {\bibfield  {journal} {\bibinfo
  {journal} {Nature Communications}\ }\textbf {\bibinfo {volume} {13}},\
  \bibinfo {pages} {5072} (\bibinfo {year} {2022})}\BibitemShut {NoStop}%
\bibitem [{\citenamefont {Ivanov}\ and\ \citenamefont
  {Kheifets}(2017)}]{Ivanov2017}%
  \BibitemOpen
  \bibfield  {author} {\bibinfo {author} {\bibfnamefont {I.~A.}\ \bibnamefont
  {Ivanov}}\ and\ \bibinfo {author} {\bibfnamefont {A.~S.}\ \bibnamefont
  {Kheifets}},\ }\bibfield  {title} {\bibinfo {title} {Angle-dependent time
  delay in two-color xuv+ir photoemission of he and ne},\ }\href
  {https://doi.org/10.1103/PhysRevA.96.013408} {\bibfield  {journal} {\bibinfo
  {journal} {Phys. Rev. A}\ }\textbf {\bibinfo {volume} {96}},\ \bibinfo
  {pages} {013408} (\bibinfo {year} {2017})}\BibitemShut {NoStop}%
\bibitem [{\citenamefont {Hockett}(2017)}]{Hockett2017}%
  \BibitemOpen
  \bibfield  {author} {\bibinfo {author} {\bibfnamefont {P.}~\bibnamefont
  {Hockett}},\ }\bibfield  {title} {\bibinfo {title} {Angle-resolved {RABBITT}:
  theory and numerics},\ }\href {https://doi.org/10.1088/1361-6455/aa7887}
  {\bibfield  {journal} {\bibinfo  {journal} {Journal of Physics B: Atomic,
  Molecular and Optical Physics}\ }\textbf {\bibinfo {volume} {50}},\ \bibinfo
  {pages} {154002} (\bibinfo {year} {2017})}\BibitemShut {NoStop}%
\bibitem [{\citenamefont {Fuchs}\ \emph {et~al.}(2020)\citenamefont {Fuchs},
  \citenamefont {Douguet}, \citenamefont {Donsa}, \citenamefont {Martin},
  \citenamefont {Burgd\"{o}rfer}, \citenamefont {Argenti}, \citenamefont
  {Cattaneo},\ and\ \citenamefont {Keller}}]{Fuchs2020}%
  \BibitemOpen
  \bibfield  {author} {\bibinfo {author} {\bibfnamefont {J.}~\bibnamefont
  {Fuchs}}, \bibinfo {author} {\bibfnamefont {N.}~\bibnamefont {Douguet}},
  \bibinfo {author} {\bibfnamefont {S.}~\bibnamefont {Donsa}}, \bibinfo
  {author} {\bibfnamefont {F.}~\bibnamefont {Martin}}, \bibinfo {author}
  {\bibfnamefont {J.}~\bibnamefont {Burgd\"{o}rfer}}, \bibinfo {author}
  {\bibfnamefont {L.}~\bibnamefont {Argenti}}, \bibinfo {author} {\bibfnamefont
  {L.}~\bibnamefont {Cattaneo}},\ and\ \bibinfo {author} {\bibfnamefont
  {U.}~\bibnamefont {Keller}},\ }\bibfield  {title} {\bibinfo {title} {Time
  delays from one-photon transitions in the continuum},\ }\href
  {https://doi.org/10.1364/OPTICA.378639} {\bibfield  {journal} {\bibinfo
  {journal} {Optica}\ }\textbf {\bibinfo {volume} {7}},\ \bibinfo {pages} {154}
  (\bibinfo {year} {2020})}\BibitemShut {NoStop}%
\bibitem [{\citenamefont {Reid}(2003)}]{Reid2003}%
  \BibitemOpen
  \bibfield  {author} {\bibinfo {author} {\bibfnamefont {K.~L.}\ \bibnamefont
  {Reid}},\ }\bibfield  {title} {\bibinfo {title} {Photoelectron angular
  distributions},\ }\href
  {https://doi.org/10.1146/annurev.physchem.54.011002.103814} {\bibfield
  {journal} {\bibinfo  {journal} {Annual Review of Physical Chemistry}\
  }\textbf {\bibinfo {volume} {54}},\ \bibinfo {pages} {397} (\bibinfo {year}
  {2003})}\BibitemShut {NoStop}%
\bibitem [{\citenamefont {Joachain}(1975)}]{Joachain1975}%
  \BibitemOpen
  \bibfield  {author} {\bibinfo {author} {\bibfnamefont {C.}~\bibnamefont
  {Joachain}},\ }\href@noop {} {\emph {\bibinfo {title} {Quantum Collision
  Theory}}}\ (\bibinfo  {publisher} {North-Holland Publishing Company},\
  \bibinfo {year} {1975})\BibitemShut {NoStop}%
\bibitem [{\citenamefont {Boll}\ \emph {et~al.}(2022)\citenamefont {Boll},
  \citenamefont {Martini},\ and\ \citenamefont {Foj\'on}}]{Boll2022a}%
  \BibitemOpen
  \bibfield  {author} {\bibinfo {author} {\bibfnamefont {D.~I.~R.}\
  \bibnamefont {Boll}}, \bibinfo {author} {\bibfnamefont {L.}~\bibnamefont
  {Martini}},\ and\ \bibinfo {author} {\bibfnamefont {O.~A.}\ \bibnamefont
  {Foj\'on}},\ }\bibfield  {title} {\bibinfo {title} {Analytical model for
  attosecond time delays and fano's propensity rules in the continuum},\ }\href
  {https://doi.org/10.1103/PhysRevA.106.023116} {\bibfield  {journal} {\bibinfo
   {journal} {Phys. Rev. A}\ }\textbf {\bibinfo {volume} {106}},\ \bibinfo
  {pages} {023116} (\bibinfo {year} {2022})}\BibitemShut {NoStop}%
\bibitem [{\citenamefont {Weisstein}()}]{HAT}%
  \BibitemOpen
  \bibfield  {author} {\bibinfo {author} {\bibfnamefont {E.~W.}\ \bibnamefont
  {Weisstein}},\ }\href
  {https://mathworld.wolfram.com/HarmonicAdditionTheorem.html} {\bibinfo
  {title} {Harmonic addition theorem}},\ \bibinfo {note} {from MathWorld--A
  Wolfram Web Resource.}\BibitemShut {Stop}%
\bibitem [{\citenamefont {Tulsky}\ and\ \citenamefont
  {Bauer}(2020)}]{Tulsky2020}%
  \BibitemOpen
  \bibfield  {author} {\bibinfo {author} {\bibfnamefont {V.}~\bibnamefont
  {Tulsky}}\ and\ \bibinfo {author} {\bibfnamefont {D.}~\bibnamefont {Bauer}},\
  }\bibfield  {title} {\bibinfo {title} {Qprop with faster calculation of
  photoelectron spectra},\ }\href
  {https://doi.org/https://doi.org/10.1016/j.cpc.2019.107098} {\bibfield
  {journal} {\bibinfo  {journal} {Computer Physics Communications}\ }\textbf
  {\bibinfo {volume} {251}},\ \bibinfo {pages} {107098} (\bibinfo {year}
  {2020})}\BibitemShut {NoStop}%
\bibitem [{\citenamefont {Jayadevan}\ and\ \citenamefont
  {Thayyullathil}(2001)}]{Jayadevan2001}%
  \BibitemOpen
  \bibfield  {author} {\bibinfo {author} {\bibfnamefont {A.~P.}\ \bibnamefont
  {Jayadevan}}\ and\ \bibinfo {author} {\bibfnamefont {R.~B.}\ \bibnamefont
  {Thayyullathil}},\ }\bibfield  {title} {\bibinfo {title} {Two-photon
  ionization of atomic hydrogen above the one-photon ionization threshold},\
  }\href {https://doi.org/10.1088/0953-4075/34/4/317} {\bibfield  {journal}
  {\bibinfo  {journal} {Journal of Physics B: Atomic, Molecular and Optical
  Physics}\ }\textbf {\bibinfo {volume} {34}},\ \bibinfo {pages} {699}
  (\bibinfo {year} {2001})}\BibitemShut {NoStop}%
\bibitem [{\citenamefont {Boll}\ \emph {et~al.}()\citenamefont {Boll},
  \citenamefont {Martini}, \citenamefont {Palacios},\ and\ \citenamefont
  {Foj\'on}}]{DataBoll2022b}%
  \BibitemOpen
  \bibfield  {author} {\bibinfo {author} {\bibfnamefont {D.~I.~R.}\
  \bibnamefont {Boll}}, \bibinfo {author} {\bibfnamefont {L.}~\bibnamefont
  {Martini}}, \bibinfo {author} {\bibfnamefont {A.}~\bibnamefont {Palacios}},\
  and\ \bibinfo {author} {\bibfnamefont {O.~A.}\ \bibnamefont {Foj\'on}},\
  }\href@noop {} {\bibinfo {title} {Angularly resolved atomic time delays}},\
  \bibinfo {howpublished}
  {\url{https://ri.conicet.gov.ar/handle/11336/182388}},\ \bibinfo {note} {from
  Repositorio Institucional CONICET Digital}\BibitemShut {NoStop}%
\bibitem [{Note1()}]{Note1}%
  \BibitemOpen
  \bibinfo {note} {The next order term in the Taylor expansion around $\Delta
  =0$ of the solutions to the quadratic equation are negligible due to $ \vert
  \Delta \vert \ll 1$}\BibitemShut {NoStop}%
\bibitem [{\citenamefont {Boll}\ \emph {et~al.}(2020)\citenamefont {Boll},
  \citenamefont {Martini}, \citenamefont {Foj\'on},\ and\ \citenamefont
  {Palacios}}]{Boll2020}%
  \BibitemOpen
  \bibfield  {author} {\bibinfo {author} {\bibfnamefont {D.~I.~R.}\
  \bibnamefont {Boll}}, \bibinfo {author} {\bibfnamefont {L.}~\bibnamefont
  {Martini}}, \bibinfo {author} {\bibfnamefont {O.~A.}\ \bibnamefont
  {Foj\'on}},\ and\ \bibinfo {author} {\bibfnamefont {A.}~\bibnamefont
  {Palacios}},\ }\bibfield  {title} {\bibinfo {title} {Off-resonance-enhanced
  polarization control in two-color atomic ionization},\ }\href
  {https://doi.org/10.1103/PhysRevA.101.013428} {\bibfield  {journal} {\bibinfo
   {journal} {Phys. Rev. A}\ }\textbf {\bibinfo {volume} {101}},\ \bibinfo
  {pages} {013428} (\bibinfo {year} {2020})}\BibitemShut {NoStop}%
\bibitem [{\citenamefont {Maquet}\ and\ \citenamefont
  {Ta\"ieb}(2007)}]{Maquet2007}%
  \BibitemOpen
  \bibfield  {author} {\bibinfo {author} {\bibfnamefont {A.}~\bibnamefont
  {Maquet}}\ and\ \bibinfo {author} {\bibfnamefont {R.}~\bibnamefont
  {Ta\"ieb}},\ }\bibfield  {title} {\bibinfo {title} {Two-colour ir+xuv
  spectroscopies: the ''soft-photon approximation''},\ }\href@noop {}
  {\bibfield  {journal} {\bibinfo  {journal} {Journal of Modern Optics}\
  }\textbf {\bibinfo {volume} {54}},\ \bibinfo {pages} {1847} (\bibinfo {year}
  {2007})}\BibitemShut {NoStop}%
\bibitem [{\citenamefont {Picard}\ \emph {et~al.}(2014)\citenamefont {Picard},
  \citenamefont {Manschwetus}, \citenamefont {G\'el\'eoc}, \citenamefont
  {B\"ottcher}, \citenamefont {Casagrande}, \citenamefont {Lin}, \citenamefont
  {Ruchon}, \citenamefont {Carr\'e}, \citenamefont {Hergott}, \citenamefont
  {Lepetit}, \citenamefont {Ta\"{\i}eb}, \citenamefont {Maquet},\ and\
  \citenamefont {Huetz}}]{Picard2014}%
  \BibitemOpen
  \bibfield  {author} {\bibinfo {author} {\bibfnamefont {Y.~J.}\ \bibnamefont
  {Picard}}, \bibinfo {author} {\bibfnamefont {B.}~\bibnamefont {Manschwetus}},
  \bibinfo {author} {\bibfnamefont {M.}~\bibnamefont {G\'el\'eoc}}, \bibinfo
  {author} {\bibfnamefont {M.}~\bibnamefont {B\"ottcher}}, \bibinfo {author}
  {\bibfnamefont {E.~M.~S.}\ \bibnamefont {Casagrande}}, \bibinfo {author}
  {\bibfnamefont {N.}~\bibnamefont {Lin}}, \bibinfo {author} {\bibfnamefont
  {T.}~\bibnamefont {Ruchon}}, \bibinfo {author} {\bibfnamefont
  {B.}~\bibnamefont {Carr\'e}}, \bibinfo {author} {\bibfnamefont {J.-F.}\
  \bibnamefont {Hergott}}, \bibinfo {author} {\bibfnamefont {F.}~\bibnamefont
  {Lepetit}}, \bibinfo {author} {\bibfnamefont {R.}~\bibnamefont {Ta\"{\i}eb}},
  \bibinfo {author} {\bibfnamefont {A.}~\bibnamefont {Maquet}},\ and\ \bibinfo
  {author} {\bibfnamefont {A.}~\bibnamefont {Huetz}},\ }\bibfield  {title}
  {\bibinfo {title} {Attosecond evolution of energy- and angle-resolved
  photoemission spectra in two-color (xuv $+$ ir) ionization of rare gases},\
  }\href {https://doi.org/10.1103/PhysRevA.89.031401} {\bibfield  {journal}
  {\bibinfo  {journal} {Phys. Rev. A}\ }\textbf {\bibinfo {volume} {89}},\
  \bibinfo {pages} {031401} (\bibinfo {year} {2014})}\BibitemShut {NoStop}%
\bibitem [{\citenamefont {Boll}\ and\ \citenamefont
  {Foj{\'{o}}n}(2016)}]{Boll2016}%
  \BibitemOpen
  \bibfield  {author} {\bibinfo {author} {\bibfnamefont {D.~I.~R.}\
  \bibnamefont {Boll}}\ and\ \bibinfo {author} {\bibfnamefont {O.~A.}\
  \bibnamefont {Foj{\'{o}}n}},\ }\bibfield  {title} {\bibinfo {title} {Atomic
  {RABBITT}-like experiments framed as diatomic molecules},\ }\href
  {https://doi.org/10.1088/0953-4075/49/18/185601} {\bibfield  {journal}
  {\bibinfo  {journal} {Journal of Physics B: Atomic, Molecular and Optical
  Physics}\ }\textbf {\bibinfo {volume} {49}},\ \bibinfo {pages} {185601}
  (\bibinfo {year} {2016})}\BibitemShut {NoStop}%
\bibitem [{\citenamefont {Boll}\ and\ \citenamefont
  {Foj{\'{o}}n}(2017)}]{Boll2017}%
  \BibitemOpen
  \bibfield  {author} {\bibinfo {author} {\bibfnamefont {D.~I.~R.}\
  \bibnamefont {Boll}}\ and\ \bibinfo {author} {\bibfnamefont {O.~A.}\
  \bibnamefont {Foj{\'{o}}n}},\ }\bibfield  {title} {\bibinfo {title}
  {Attosecond polarization control in atomic {RABBITT}-like experiments
  assisted by a circularly polarized laser},\ }\href
  {https://doi.org/10.1088/1361-6455/aa8c9f} {\bibfield  {journal} {\bibinfo
  {journal} {Journal of Physics B: Atomic, Molecular and Optical Physics}\
  }\textbf {\bibinfo {volume} {50}},\ \bibinfo {pages} {235604} (\bibinfo
  {year} {2017})}\BibitemShut {NoStop}%
\end{thebibliography}%

\end{document}